\documentclass[review]{elsarticle} 
\usepackage{geometry}%
\usepackage{setspace}

\usepackage{amssymb}
\usepackage{lipsum}

\usepackage{color,array,amsthm}
\usepackage{epstopdf}
\usepackage{graphicx}
\usepackage[T1]{fontenc}
\usepackage{mathptmx}
\fontfamily{ptm}\selectfont 
\usepackage{tabularx}

\usepackage{bm}

\usepackage{booktabs}

\usepackage{amsmath} 

\usepackage{hyperref}
\usepackage{cleveref}
\fontfamily{cmr}\selectfont 

\begin{document}
\begin{frontmatter}


\linespread{2}
\title{Adaptive Model Predictive Control for Engine-Driven Ducted Fan Lift Systems using an Associated Linear Parameter Varying Model }


\author[a]{Hanjie Jiang}

\author[a]{Ye Zhou \corref{cor}}
\ead{zhouye@usm.my}

\author[a]{Hann Woei Ho}
\author[a]{Wenjie Hu}

\cortext[cor]{Corresponding author}

\affiliation[a]{
  organization = {School of Aerospace Engineering, Engineering Campus, Universiti Sains Malaysia},
  city         = {Nibong Tebal},
  postcode     = {14300},
  country      = {Malaysia},
}

\begin{abstract}
\indent
Ducted fan lift systems (DFLSs) powered by two-stroke aviation piston engines present a challenging control problem due to their complex multivariable dynamics. Current controllers for these systems typically rely on proportional-integral algorithms combined with data tables, which rely on accurate models and are not adaptive to handle time-varying dynamics or system uncertainties. 
This paper proposes a novel adaptive model predictive control (AMPC) strategy with an associated linear parameter varying (LPV) model for controlling the engine-driven DFLS. 
This LPV model is derived from a global network model, which is trained off-line with data obtained from a general mean value engine model for two-stroke aviation engines.
Different network models, including multi-layer perceptron, Elman, and radial basis function (RBF), are evaluated and compared in this study. The results demonstrate that the RBF model exhibits higher prediction accuracy and robustness in the DFLS application. 
Based on the trained RBF model, the proposed AMPC approach constructs an associated network that directly outputs the LPV model parameters as an adaptive, robust, and efficient prediction model. The efficiency of the proposed approach is demonstrated through numerical simulations of a vertical take-off thrust preparation process for the DFLS. 
The simulation results indicate that the proposed AMPC method can effectively control the DFLS thrust with a relative error below 3.5\%.
\end{abstract}




\end{frontmatter}




\section{Introduction}
\label{sec:introduction}
The ducted fan lift system (DFLS) is widely used in current vertical takeoff and landing (VTOL) aircraft, many of which use fuel engines as the ducted fan drive devices\cite{2020Boeing, 2015A,2021Martin, Frank2016DARPA}. Compared to an electrically powered DFLS, a fuel-engine-powered DFLS is more likely to meet the requirements of high power and high energy density, thereby delivering superior flight performance. Two-stroke aviation piston engines, which serve as the power unit of the DFLS, have rapid, highly nonlinear dynamics with state and input constraints \cite{2000motive, Yixuan2019Efficiency}. In addition, the complex geometry of the ducted fan of the DFLS makes it difficult to analyze its aerodynamic properties \cite{2020Aerodynamic, 2010article}. These characteristics make the engine-driven DFLS a multivariable system with tightly coupled nonlinear dynamics, posing modeling and control challenges. 

Many current spark ignition (SI) engines employ feed-forward control based on a state observer and a proportional-integral (PI) type feedback control\cite{2010Model-based,2002Air}. Typically, look-up tables are used to implement the PI controller, which necessitates a laborious process of calibration and tuning. When the state of the engine changes rapidly, control accuracy tends to decrease\cite{2002Air}. 
To address these challenges, researchers have developed advanced control strategies \cite{zhou2021, 1999towards, 2008Adaptive} that can be applied to SI engines, enabling more precise and energy-efficient control. 
For instance, a global optimal control method based on $H\infty$ theory was proposed for systematic control of air-fuel ratio (AFR) with high robustness and quick responses \cite{1999towards}. Experimental results applied to a four-cylinder multi-port injection (MPI) engine indicate that the AFR control error can be limited to within 3\% across a broad spectrum of operating conditions. However, the sixth-order controller is unsuitable for real-time computation, restricting its practical applications. 
Another approach applied the dynamical sliding mode control (SMC) with a radial basis function (RBF) neural network model to a piston engine \cite{2008Adaptive}. This work demonstrated that the SMC algorithm is robust, fast, and insensitive to parameter changes and external disturbances within the context of nonlinear system control problems. However, it's important to acknowledge that as the system state approaches the sliding mode surface, achieving precise sliding along the surface to reach the equilibrium point becomes challenging, resulting in non-convergence. 

In contrast, Model predictive control (MPC) is widely recognized as an advanced control method in practical control engineering\cite{YU2022246,2021MPC}. It can effectively address multivariable constrained optimal control problems and offers the advantages of simplicity, straightforward design, high stability, robustness, and adaptability\cite{2013book, 2022Stochastic, 2018Fast}. MPC has proven to be effective for fuel engine systems containing multiple variables, nonlinear dynamics, and time delays\cite{LI2018714,2010Online}. 
However, the dynamics of the DFLS can be highly nonlinear, particularly concerning the throttle position (TPS) and the injection fuel mass flow. It may even exhibit open-loop instability during the transition between operation points. Thus, the performance of a linear MPC designed for a specific operating condition will degrade near another operating point. 
Recent studies\cite{GAO2006323, ILKA2015912, CARVALHO202157} explored gain-scheduled MPC to cope with nonlinear systems when the plant models have different orders or time delays.  This approach involves incorporating multiple predictive controllers into a gain-scheduled MPC for various operating points, switchable based on a predetermined scheduling policy. However, gain-scheduled MPC faces limitations in ensuring control accuracy during transitions for nonlinear systems and demands substantial computing resources \cite{GAO2006323}. 
In the realm of SI engine control \cite{Kazuyoshi0Nonlinear, LI2018714}, nonlinear MPC has gained traction due to its potential for more precise control performance. Yet, achieving effective and stable control performance necessitates meticulous considerations, such as computational complexity, convergence challenges, and parameter selection, particularly in real-time or large-scale applications.

In piston engine control, adaptive Model Predictive Control (AMPC) has emerged as an alternative, utilizing an online-identified linear model through successive linearization or online model estimation. This approach offers enhanced efficiency compared to nonlinear MPC\cite{Yu2017Adaptive, 2008Adaptive, 2009Neural, 2010Online}. The successive linearization method employs a set of nonlinear ordinary differential and algebraic equations to build the plant model and derives the linear time-invariant (LTI) approximation at the current operating condition to update the model parameters. However, when dealing with highly nonlinear systems, successive linearization techniques typically require a large number of iterations and computational resources. 
Online AMPC, on the other hand, has good control accuracy and robustness, but it still comes at the cost of significantly increased control optimization time and memory requirements due to extensive online operations. During the data sample identification process, online AMPC is susceptible to model errors stemming from issues such as delays, noise, and insufficient excitation.

The objective of this paper is to develop a prediction model for AMPC that mitigates the need for complex and time-consuming online operations. 
Our proposed approach involves the direct construction of an associated linear parameter varying (LPV) model, derived from a nonlinear global model. To achieve this,
the global model will be approximated using neural networks, chosen for their high approximation capability and successful applications in modeling for predictive engine control\cite{2009Neural, 2010Online, 2020Compression}. 
This paper investigated and compared the performance of three different network models: the multi-layer perceptron (MLP) \cite{1996Efficient}, radial basis function (RBF) \cite{1995Radial}, and the Elman network \cite{2006Elman}. Based on the experimental results, the RBF network emerges as the most suitable choice for representing the SI engine in the context of DFLS AMPC. 
By deriving an associated linear parameter varying (LPV) model from the global network model, we depart from the conventional practice of employing the nonlinear model network as a prediction model or identifying a linear model online. This shift reduces runtime computational demands and memory usage while simultaneously bolstering DFLS control robustness.  
In this paper, the network model undergoes training using the input and output data generated by a mean value engine model (MVEM)\cite{2022M}, and the ducted fan model of the DFLS will be derived using the method of theoretical design and evaluation\cite{2022Aerodynamic}. This study focuses on the basic thrust control of the DFLS, while the thrust fine-tuning and direction are controlled by the exit louvers configured at the outlets of the duct\cite{1972Fan}. Specifically, the control strategy of the DFLS optimizes and constrains the working state of the engine by establishing a desired thrust baseline engaged by the engine output power and a desired AFR. In addition, based on the multistep-ahead prediction of both the aerodynamic thrust and the AFR, the optimal control to track and maintain the desired value is obtained as the engine state changes.

This paper presents the following \textbf{main contributions}: 
1) This study investigates and compares the MLP, Elman, and RBF network models, and finds that the RBF model is more accurate in its predictions and more robust in the DFLS application. 
2) The study proposes an innovative AMPC method employing a novel associated LPV model directly derived from an off-line trained RBF global model, to efficiently update the prediction model state.
3) The proposed RBF model-based AMPC is applied to the nonlinear control of a DFLS, demonstrating the practicality of the method.

The remainder of the paper is structured as follows: Section 2 introduces the DFLS dynamic model consisting of an MVEM and a theoretical ducted fan model. In Section 3, three types of engine neural network models are established, trained, and compared. Section 4 proposes the AMPC controller with an LPV model generated from the model network trained in Section 3. Section 5 concludes the paper and outlines future research directions.

\section{The ducted fan lift system}
\begin{figure}[h]
\centering                
\includegraphics[width=0.55\linewidth]{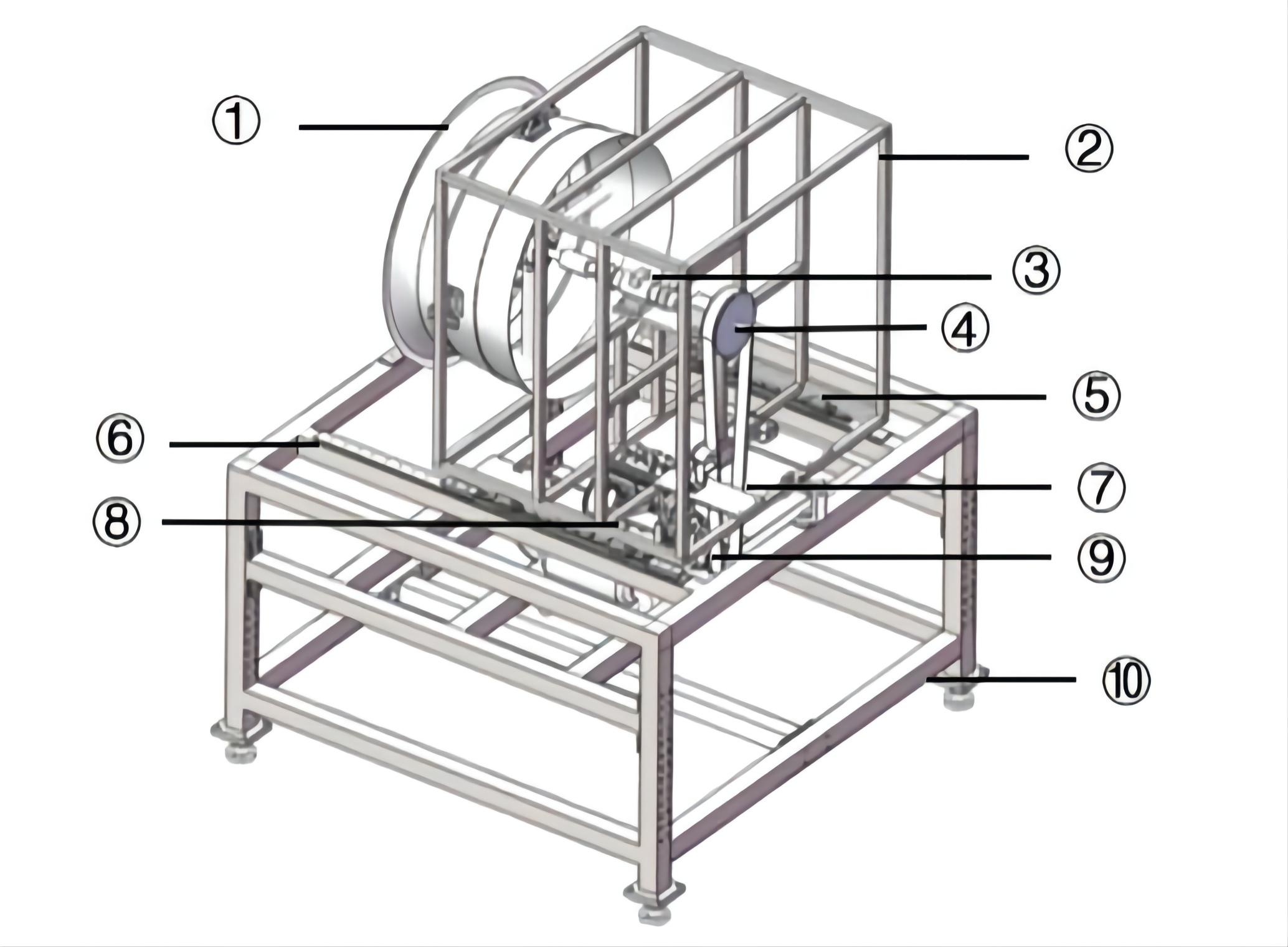} 
\caption{The DFLS control test bench scheme (1: ducted fan, 2: installation rack for the ducted fan, 3: torque and rotational speed measuring unit, 4: belt pulley, 5: forcemeter, 6: sliding rail, 7: pulley belt, 8: SI engine, 9: belt pulley, 10: test bench base).}\label{fig:1}
\end{figure}

The DFLS dynamic model is comprised of the SI engine and ducted fan modules. As the propulsion system, the SI engine utilizes the transmission mechanism to finally rotate the fan shaft and generate lift force. Figure\ref{fig:1} depicts the control test bench scheme of the DFLS demonstrator, illustrating the assembly relationship between the components.

\subsection{The engine's dynamics}
For the development of engine controllers and their validation with simulations, mathematical modeling of engine dynamics is essential\cite{2012Dynamic}. The MVEM is a widely used mathematical engine model that has achieved many successes in real-time simulation and control of automobile and ship engines\cite{2012Mean,2018Development,2017Simulation}. The general MVEM uses empirical equations to construct engine sub-models, which reduces engine modeling time and computational cost by a significant margin. The implementation of the MVEM combines quasi-static and volumetric models, dividing the two-stroke engine into four independent volumetric control units\cite{2022M}: the dynamics of intake manifold, the crankcase and cylinder module, the dynamics of fuel injection, and the dynamics of the crankshaft.

The equation for the crankshaft speed state in the general MVEM is\cite{2022M}
\begin{equation} \label{eq:1}
\Dot\omega = \frac{{H}_{u}\eta_i(1-k_f)\Dot{m}_{f}(t-\tau_d)}{I\omega}-\frac{{P}_{f}+{P}_{b}}{I\omega},
\end{equation}
where $\Dot\omega$ is the angular acceleration of the crankshaft, $H_{u}$ is the lower heating value of the fuel, $\eta_i$ is the thermal efficiency, $t$ is the time, $\tau_d$ is the injection-torque time delay, and $k_f$ is the proportionality coefficient constant for fuel loss resulting from short-circuiting and overflow losses during the scavenging process. $I$ is the inertia of the engine, $P_f$ is the friction loss, and $P_b$ is the load power of the engine.  The torque of the engine can be expressed as\cite{2022M}
\begin{equation} \label{eq:2}
{Q}_{eng} = \frac{{H}_{u}\eta_i(1-k_f)\Dot{m}_{f}(t-\tau_d)}{\omega}-\frac{{P}_{f}}{\omega}.
\end{equation} 
The intake airflow is mixed with fuel during the engine's intake process, and the normalized air-fuel ratio is
\begin{equation} \label{eq:3}
\lambda = {\frac{\Dot{m}_{as}}{\Dot{m}_{f}L_{th}}},
\end{equation}
where $L_{th}$ is the stoichiometric air-fuel ratio.

\subsection{Ducted fan dynamics}
Due to the complexity of the geometry, it is difficult to analyze the aerodynamic properties of ducted fans. This paper adopts a modeling method for the DFLS based on the blade element theory (BET) and the momentum theory\cite{2022Aerodynamic}. According to the BET, propellers are made up of minuscule elements in the shape of airfoils along the radius of each blade\cite{2006Performance}. The resulting velocity for each element can be decomposed into rotational and translational components. The resulting aerodynamic force can be decomposed into the drag and lift. When decomposed in the plane of rotation, it produces thrust and the torque-producing force.

The total thrust $T_{UDF}$ can be expressed as\cite{2022Aerodynamic}
\begin{equation} \label{eq:4}
T_{UDF} = {\frac{1}{2}}{\rho}{V_{trans}}^2B\int_{0}^{R}{T_{c}\cdot{dr}},
\end{equation}
where $\rho$ is  the air density, $V_{trans}$ is the translational speed component, $B$ is the blade number corrective factor, $R$ is the blade radius, $T_c$ is the thrust coefficient and $dr$ represents the infinitesimal airfoils along the blade radius. Similarly, the total torque $Q_{UDF}$ can be expressed as:
\begin{equation} \label{eq:5}
Q_{UDF} = {\frac{1}{2}}{\rho}{V_{trans}}^2B\int_{0}^{R}{Q_{c}\cdot{dr}},
\end{equation}
where $Q_c$ is the torque coefficient.
The rate of energy supplied by the engine matches its power $P_{UDF}$:
\begin{equation} \label{eq:6}
P_{UDF} = 2\pi{n}{Q_{UDF}}.
\end{equation}

Using the momentum theorem, the static thrust ratio between the ducted fan and the unducted fan with the same absorbed power\cite{1948Static} can be calculated as:
\begin{equation} 
\label{eq:7}
{\frac{T_{DF}}{T_{UDF}}} = 1.26\Bigg({\frac{S_{3}}{S_{2}}}\Bigg)^{\frac{1}{3}},
\end{equation}
where $S_2$ is the area of the fan disc and $S_3$ is the area of the duct outlet. Equations \ref{eq:1}, \ref{eq:2}, \ref{eq:4}-\ref{eq:7} can be used to calculate the thrust of the ducted fan that corresponds to the engine output in this instance. Specifically, the absorbed power of the ducted fan is determined by calculating the engine output power from the MVEM output. BET is then used to calculate the thrust of an unducted fan with the same absorption power. 

Previous studies have validated the dynamic models of the DFLS\cite{2022M,2022Aerodynamic}. Both models are explicit mathematical representations, allowing for flexible combination and application. In this study, the established DFLS dynamic model will be utilized as the plant for control research, and the MVEM data will be employed to create a neural network engine model intended for the network-based controller in the following section. 

\section{Neural network model of the engine}
\begin{figure}[h]
\centering                             
\includegraphics[width=0.45\linewidth]{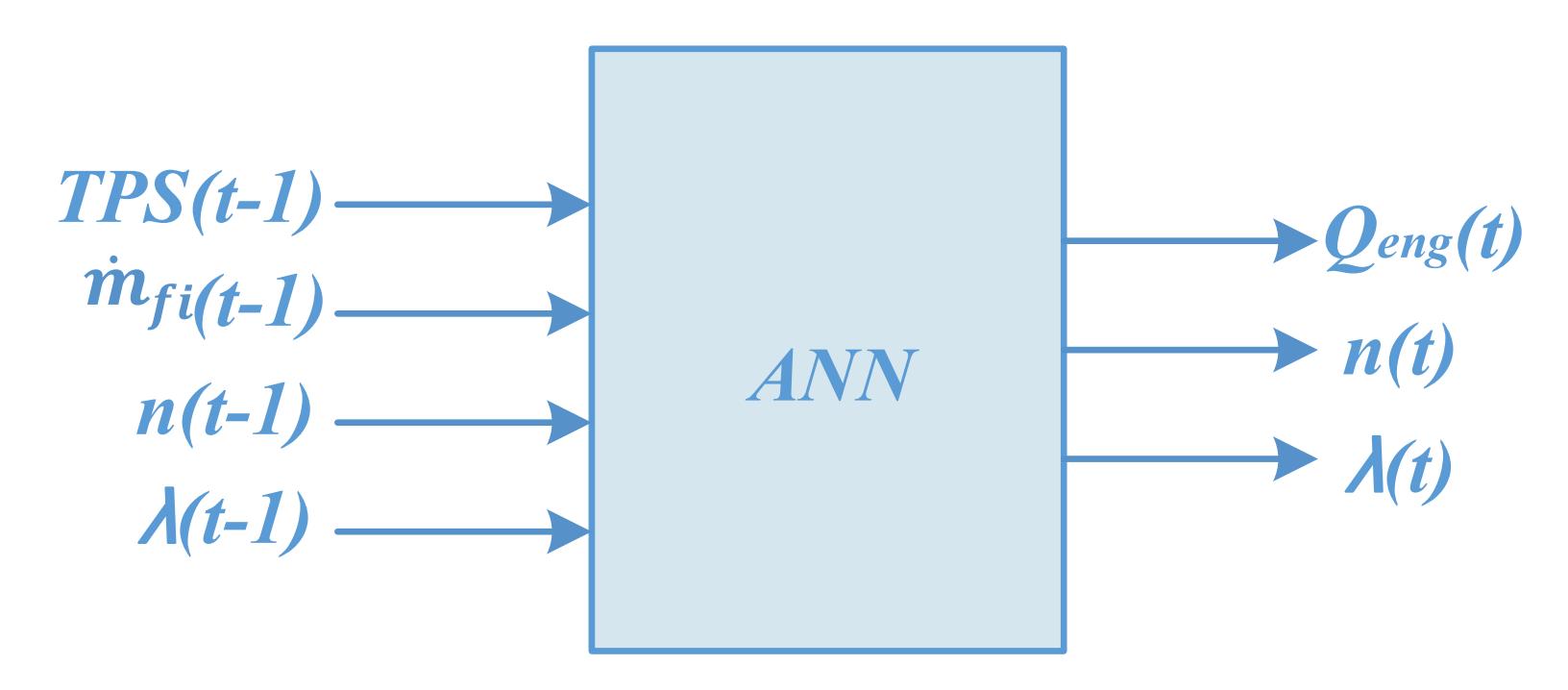} 
\caption{Expanded engine model structure}\label{fig:4}
\end{figure}

As stated previously, the engine network model is established using MLP, Elman, and RBF neural networks in this study. After comparing the accuracy and robustness of these three network models, the best one will be utilized in the development of the AMPC controller. Figure \ref{fig:4} depicts the expanded engine model, which has four inputs, fuel injection rate $\dot{m}_{fi}$, throttle position $TPS$, engine speed $n$, and normalized AFR $\lambda$, and generates three outputs, normalized AFR $\lambda$, engine speed $n$, and engine output torque $Q_{eng}$ at next instance. The engine network model developed will not be employed as an adaptive prediction model for the MPC directly, but rather to facilitate the development of an LPV prediction model generated from the trained network parameters.

\subsection{MLP network model}
\begin{figure}[h]
\centering                             
\includegraphics[width=0.65\linewidth]{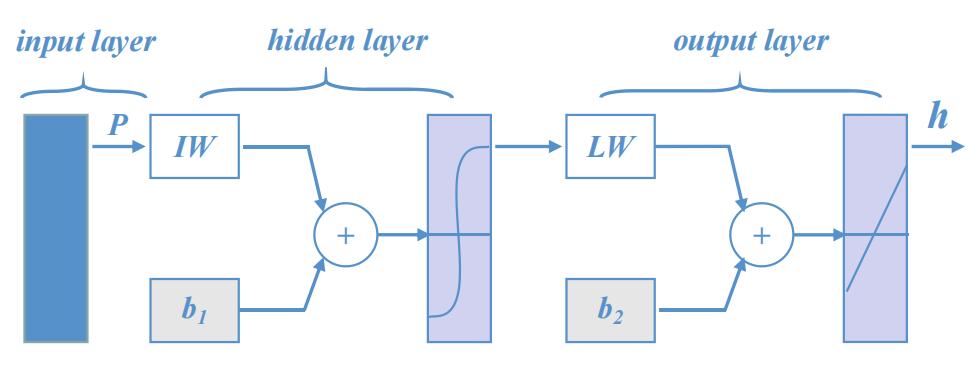} 
\caption{The concise structure of the MLP network.}\label{fig:7}
\end{figure}

Based on the gradient descent algorithm, the MLP network is a supervised learning technique\cite{1995Neural}. The MLP network structure is depicted in Figure \ref{fig:7} and the mathematical expression of the MLP network output is

\begin{equation} \label{eq:10}
\bm{h} = f_2(LW \cdot f_1(IW \cdot \bm{p}^T+\bm{b}_1)+\bm{b}_2),
\end{equation}
where $\bm{p}$ represents the input vector, $IW$ represents the weight matrix from the input layer to the hidden layer, and $LW$ represents the weight matrix from the hidden layer to the output layer. The bias vectors for the hidden layer and output layer are denoted by $\bm{b}_1$ and $\bm{b}_2$, respectively. 

The primary learning phases of MLP networks are mode forward transmission process and error back propagation (BP). In the back propagation process, when the output does not achieve the desired value, the error signal transfers back along the original path. Adjustment of general functions based on cumulative value and domain is expressed as\cite{2021Evaluation}:

\begin{equation} \label{eq:11}
W(k+1) = W(k) - \alpha \frac{\partial E_k}{\partial W(k)},
\end{equation}
\begin{equation} \label{eq:12}
\bm{b}(k+1) = \bm{b}(k) - \beta \frac{\partial E_k}{\partial b(k)},
\end{equation}
where $W(k+ 1)$ and $W(k)$ are the weights at times $k+1$ and $k$ respectively, and $\bm{b}(k + 1)$ and $\bm{b}(k)$ are the biases at times $k+1$ and $k$ respectively. The error function is denoted by $E_k$, and $\alpha$ and $\beta$ are the learning rates.

For the MLP network, different number of hidden layer nodes have been used in training experiments and a structure with a single hidden layer containing 26 nodes is chosen as shown in Figure \ref{fig:4}, which gives the minimum prediction error. The activation function of the hidden layer $f_1$ is hyperbolic tangent,and the activation function of the output layer $f_2$ is linear transfers. The learning rates $\alpha$ and $\beta$ are set as 0.1 in this MLP network to ensure the convergence of network training. In this paper, the training accuracy target is set to 0.0001. 

\subsection{Elman network model}
\begin{figure}[h]
\centering                             
\includegraphics[width=0.65\linewidth]{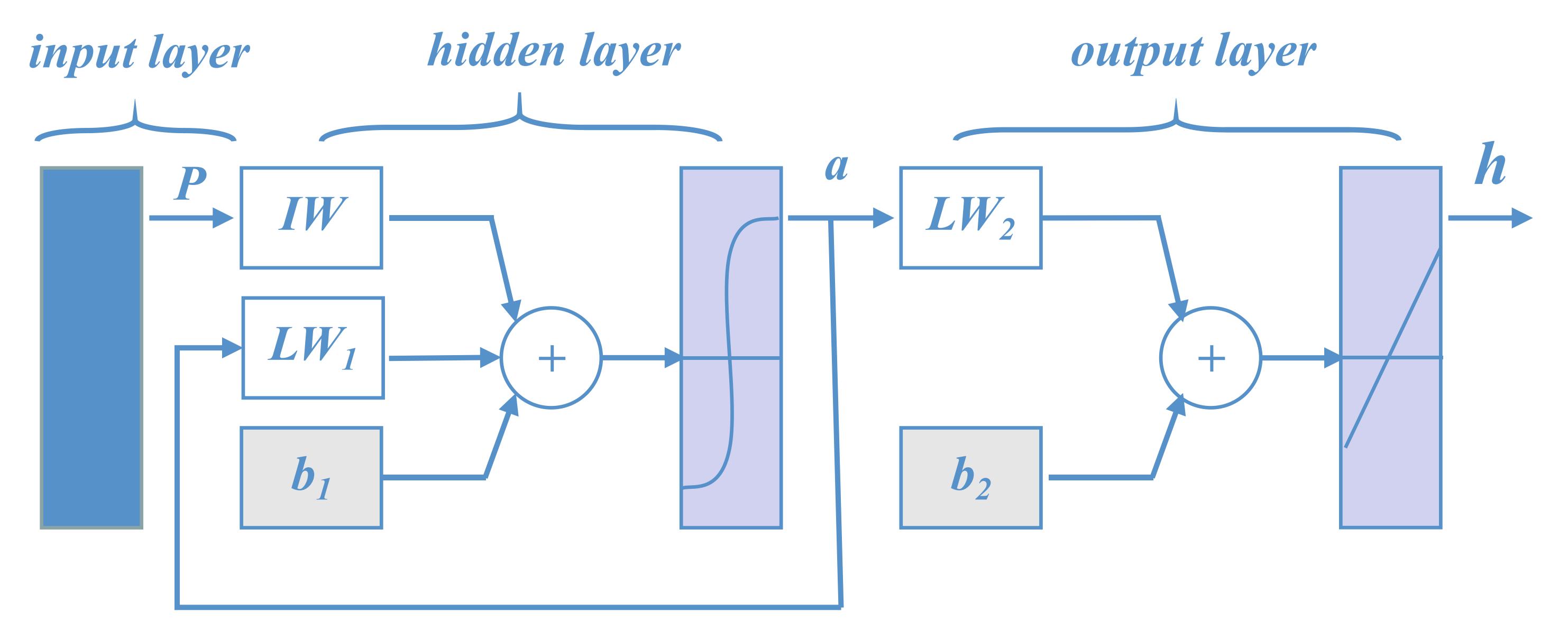} 
\caption{The structure of the Elman network.}\label{fig:8}
\end{figure}

Elman network is a well-known partial recurrent network, which lies between a traditional feed-forward perception network and a pure recurrent network\cite{2006Elman}. As shown in Figure \ref{fig:8}, a back-forward loop is included in an Elman network that is sensitive to input data history\cite{2009A}.

The Elman network is also trained using the dynamic BP algorithm, and its hidden layer activation function $f_1$ utilizes the hyperbolic tangent function, and the output layer activation function $f_2$ utilizes the linear transfer function. Due to the additional backforward loop, the Elman network's input-output relations differ from those of the MLP network, as illustrated below:
\begin{equation} \label{eq:13}
\textbf{a}(k) = f_1(IW \cdot \textbf{p} +LW_1 \cdot \textbf{a}(k-1)+\textbf{b}_1), 
\end{equation}
\begin{equation} \label{eq:14}
\textbf{h}(k) = f_2(LW_2 \cdot \textbf{a}(k)+\textbf{b}_2). 
\end{equation}

The Elman network selects a structure with a single hidden layer which contains 12 nodes, after the training experiments with different number of hidden layer nodes. The learning rate is set to be 0.01, and the training accuracy target is set to 0.0001. The maximum number of training epochs is set as 1000.

\subsection{RBF network model}
\begin{figure}[h]
\centering                             
\includegraphics[width=0.65\linewidth]{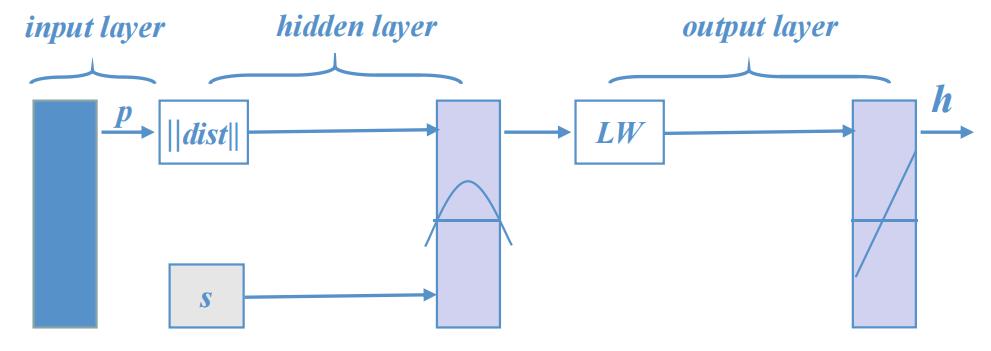} 
\caption{The concise structure of the RBF network.}\label{fig:9}
\end{figure}

RBF neural networks consist of an input layer, a single hidden layer with a radial basis activation function, and a linear output layer\cite{2009A}. Gaussian is the most frequently used activation function: 
\begin{equation} \label{eq:15.1}
\phi(\textbf{\textit{p}},\textbf{\textit{c}}) =e^{-{\frac{\|\textbf{\textit{p}}-\textbf{\textit{c}}\|^2}{s^2}}},
\end{equation}
where $\bm{c}$ is the center of the Gaussian function and $s$ is the radius, which gives a measure of the spread of the Gaussian curve. Figure \ref{fig:9} depicts the architecture of the RBF network. The distance between the input vector $\bm{p}$ and the center vector $\bm{c}$ is denoted by $\|dist\|$, and the RBF network output is:

\begin{equation} \label{eq:15}
\bm{h} = LW \cdot e^{-{\frac{\|\textbf{\textit{p}}-\textbf{\textit{c}}\|^2}{s^2}}}.
\end{equation}

The learning process of the RBF network can be divided into three stages: in the first stage, based on the distribution of the input sample, the centers and radius values of each node in the hidden layer are determined. In the second stage, the output layer weights are calculated using least-square methods. Thirdly, the parameters of the hidden layer and output layer are simultaneously adjusted based on the sample signal to improve the network precision. Using the Poggio\cite{1989A} method and the r-nearest neighborhood heuristic, the authors determined the centers $c$ and the radius $s$ of the hidden layer nodes in the RBF network. For training the network weights, the least-mean-square (LMS) algorithm is used. In this study, various numbers of hidden layer nodes have been tested, and a hidden layer with 25 centers is selected for the RBF network.
\subsection{Training and comparison of neural network models}
In the engine data collection stage, training data must be representative of typical plant behaviors in order to evaluate the performance of various engine models under practical driving conditions. Consequently, the sampled data should adequately represent the state space of the system to be controlled. The time scale of the MVEM is just sufficient to accurately describe the changes in the engine variables that change the most rapidly, which is advantageous for engine control applications \cite{1990mvem}. The trained neural network model will therefore have adequate transient and steady-state performance.

\begin{figure}[htbp]
{
    \begin{minipage}[t]{0.5\linewidth}
        \includegraphics[scale=0.42]{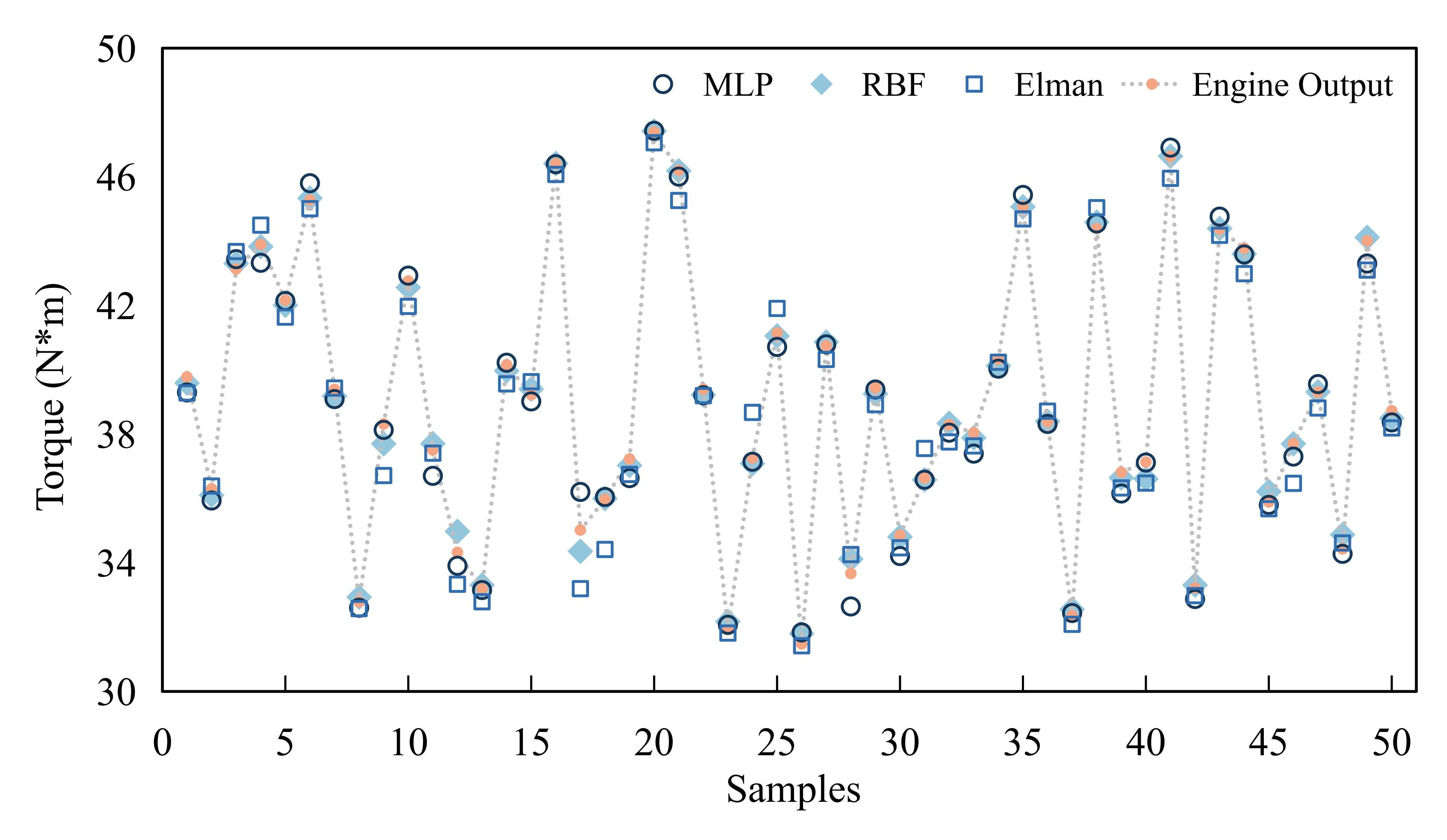} \\{\fontsize{8pt}{1em} \selectfont a) Prediction results from network models \par} 
    \end{minipage}
}
{
 	\begin{minipage}[t]{0.5\linewidth}
        \includegraphics[scale=0.431]{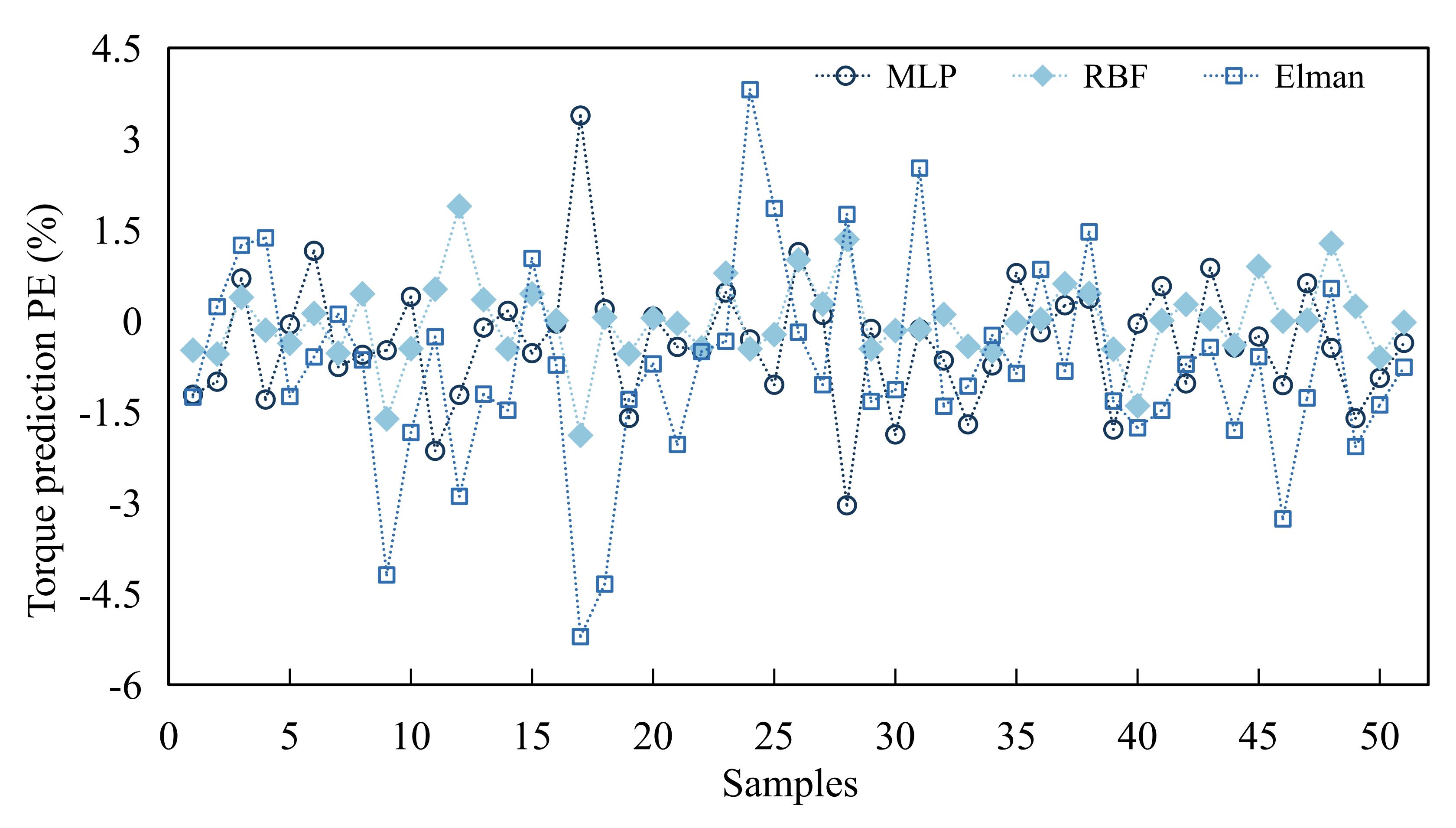} \\{\fontsize{8pt}{1em} \selectfont b) PE of prediction results \par}
    \end{minipage}
}
\caption{Comparison of predicted engine torque.}\label{fig:13}
\end{figure}

\begin{figure}[htbp]
{
    \begin{minipage}[t]{0.5\linewidth}
        \includegraphics[scale=0.42]{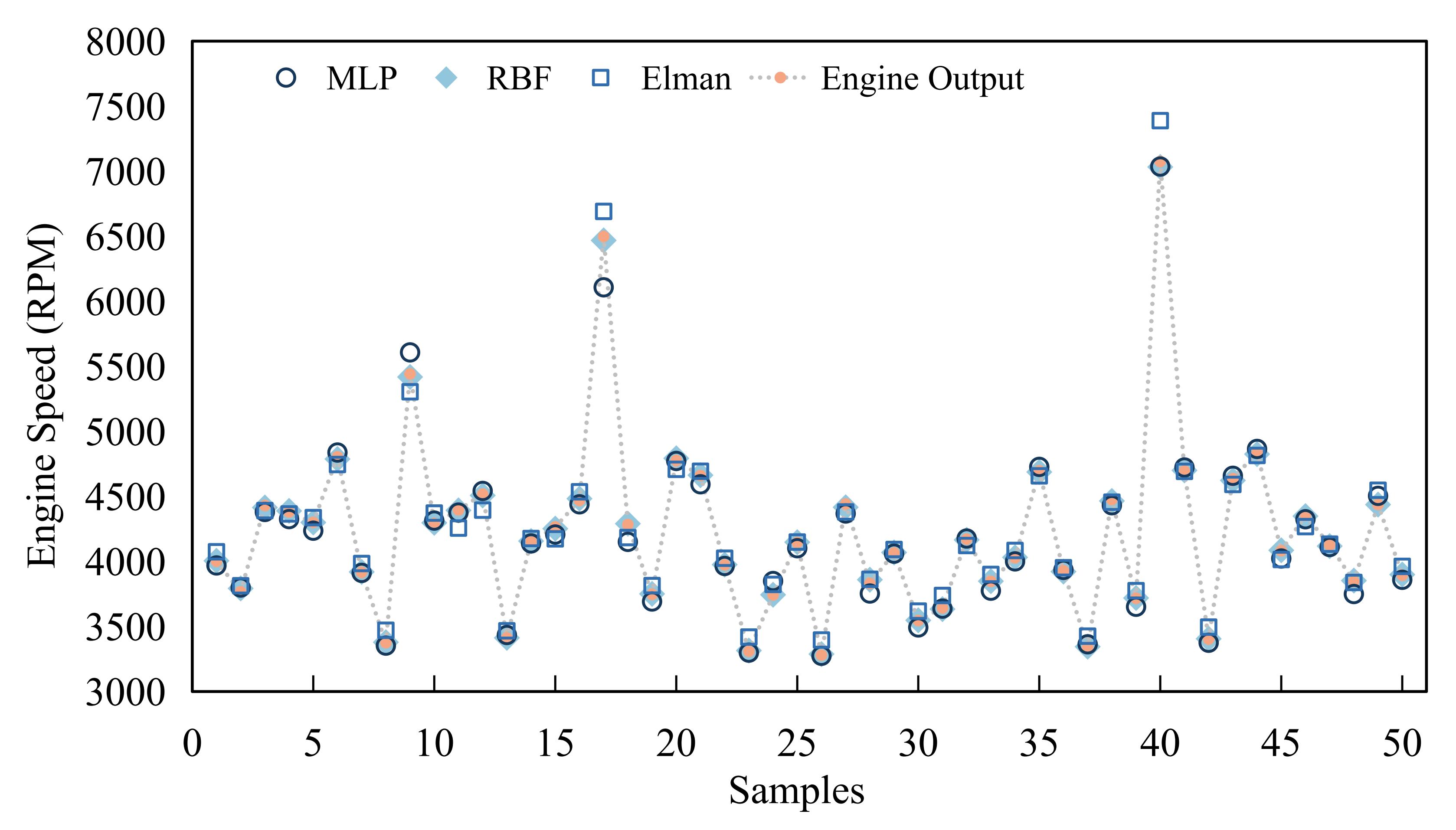} \\{\fontsize{8pt}{1em} \selectfont a) Prediction results from network models \par} 
    \end{minipage}
}
{
 	\begin{minipage}[t]{0.5\linewidth}
        \includegraphics[scale=0.43]{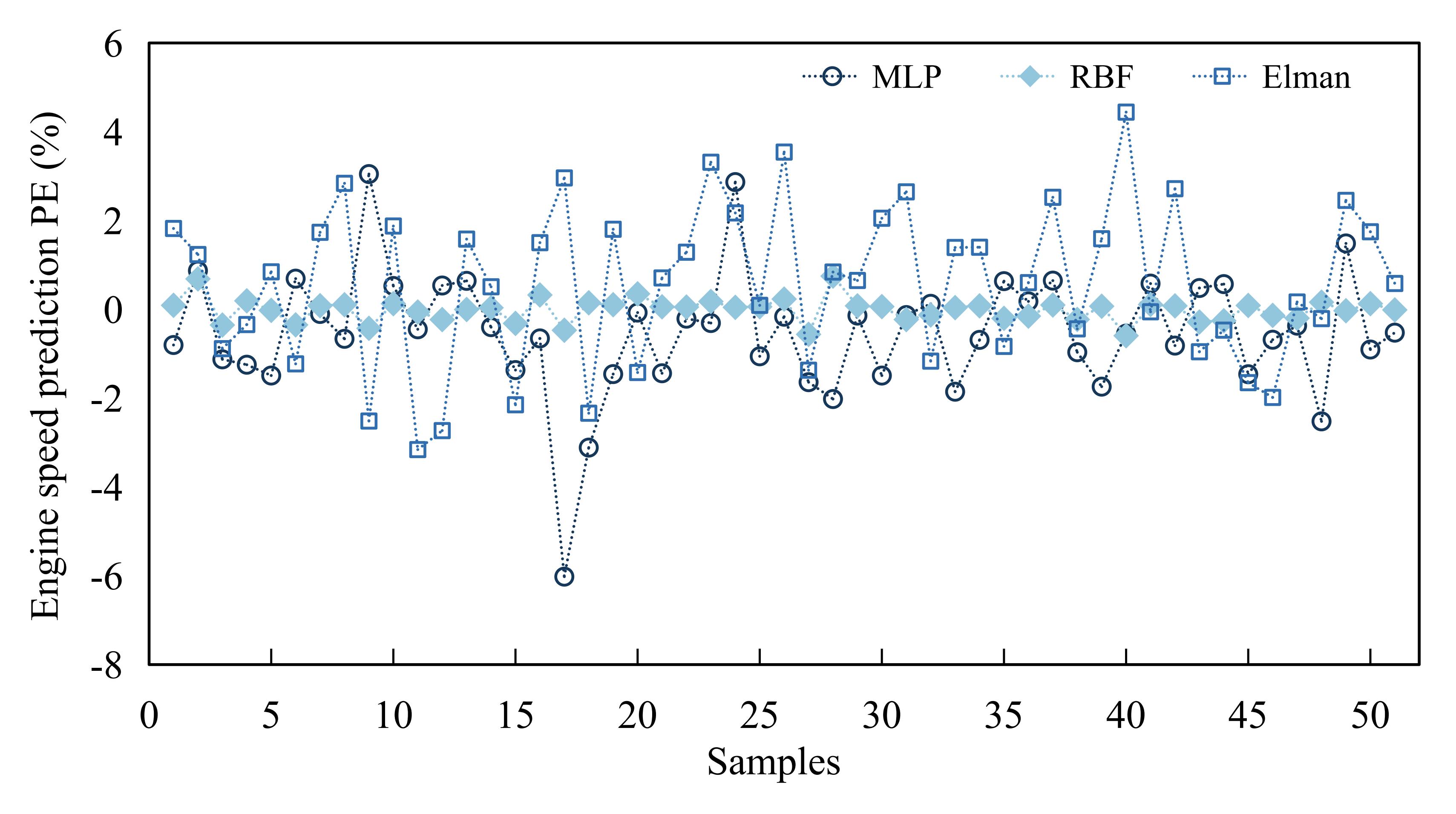} \\{\fontsize{8pt}{1em} \selectfont b) PE of prediction results \par}
    \end{minipage}
}
\caption{Comparison of predicted engine speed.}\label{fig:14}
\end{figure}

\begin{figure}[htbp]
{
    \begin{minipage}[t]{0.5\linewidth}
        \includegraphics[scale=0.43]{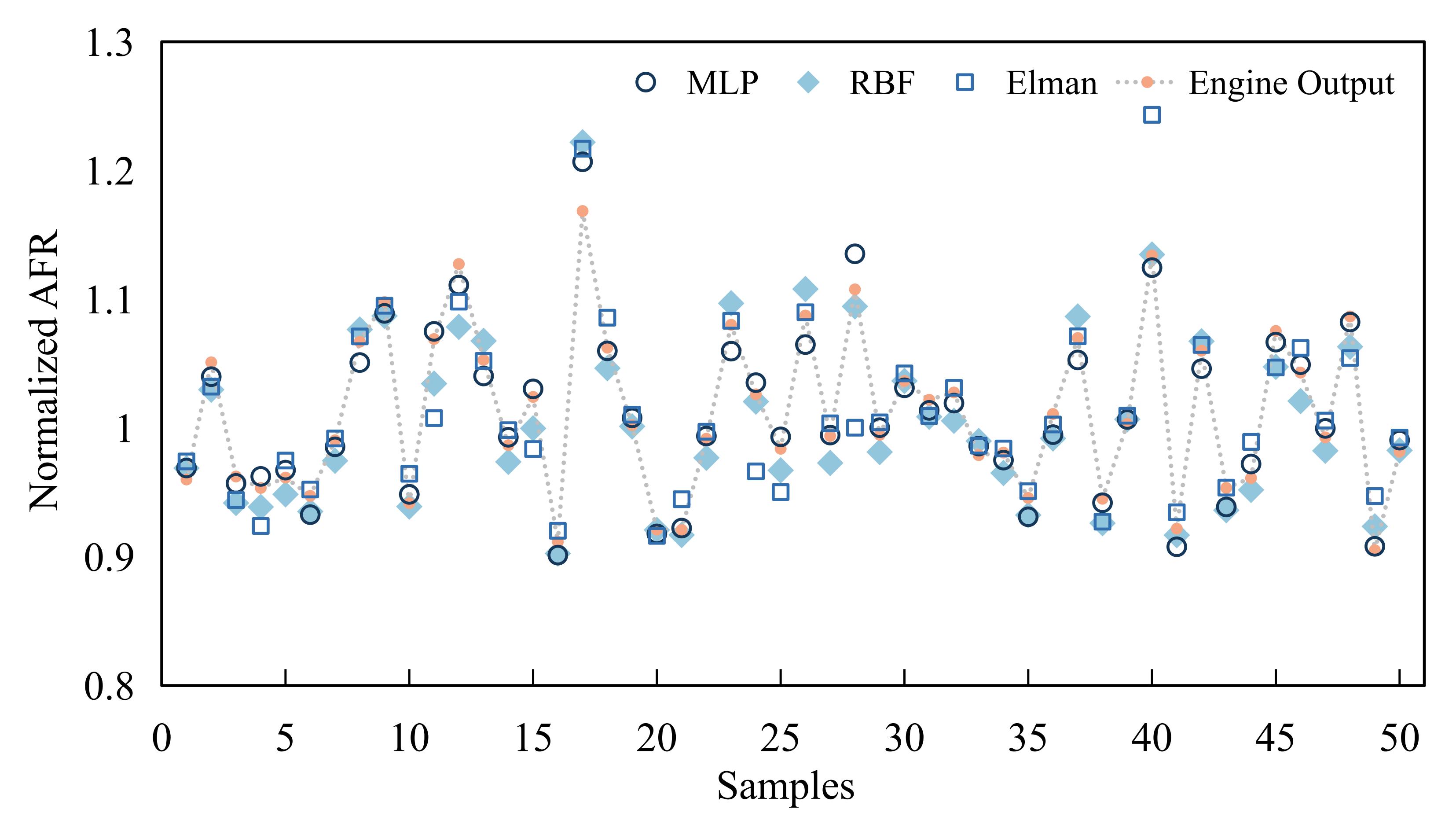} \\{\fontsize{8pt}{1em} \selectfont a) Prediction results from network models  \par}
    \end{minipage}
}
{
 	\begin{minipage}[t]{0.5\linewidth}
        \includegraphics[scale=0.421]{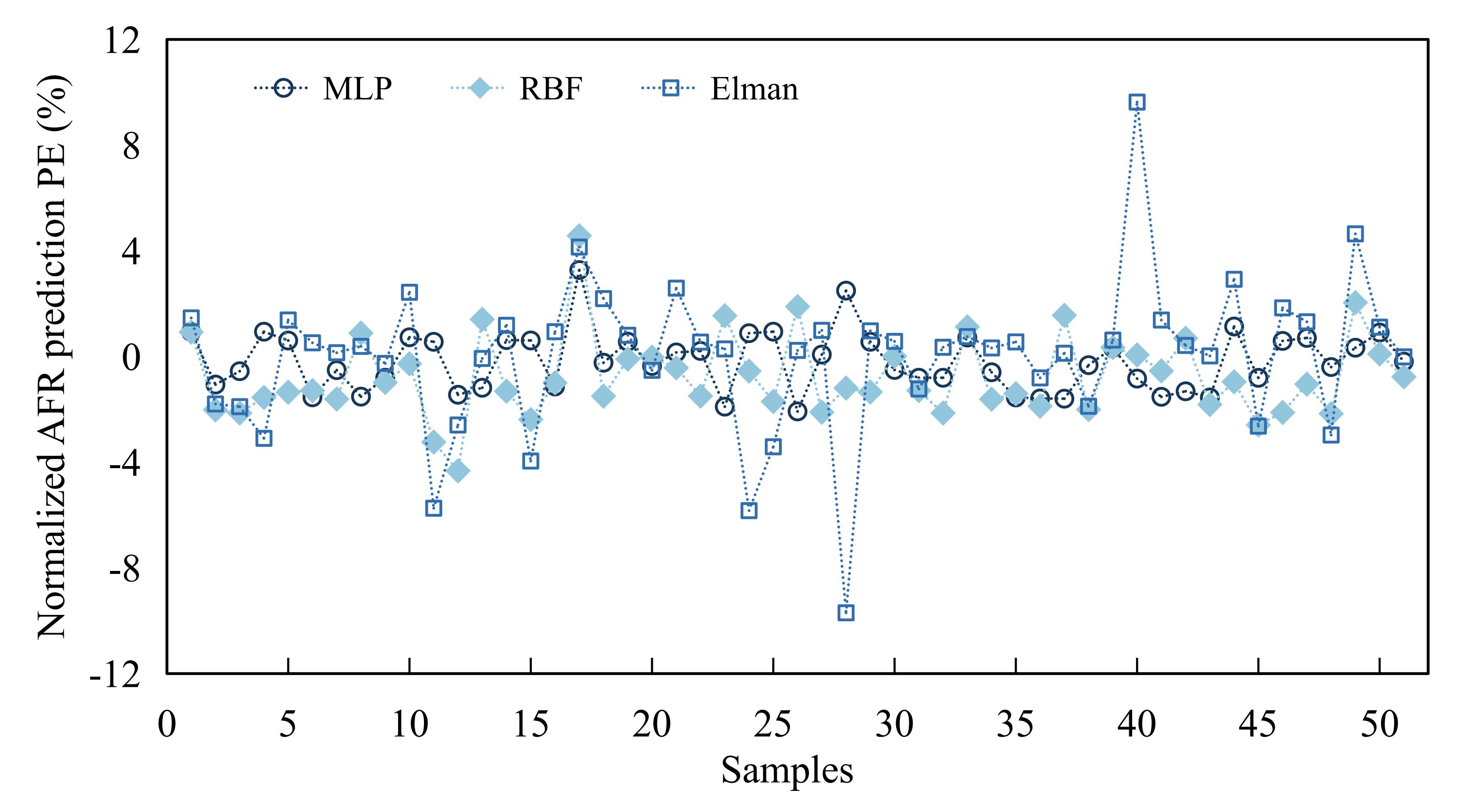} \\{\fontsize{8pt}{1em} \selectfont b) PE of prediction results \par}
    \end{minipage}
}
\caption{Comparison of predicted engine AFR.}\label{fig:15}
\end{figure}

A total of 1000 data samples are generated and separated into two groups: 950 samples for training and 50 samples for validation. Before the training and validation, all input and output data were normalized to the range of $[-1, 1]$. The Gaussian white noise (GWN) is added to the neural network training samples and its signal-to-noise ratio (SNR) is 5 db. Figures \ref{fig:13}-\ref{fig:15} compare the prediction results and the corresponding proportional error (PE) over the validation data after training with 950 samples. 
The throttle position is constrained between 5\% and 90\%, and the fuel injection rate is between 0.0011 and 0.0055 kg/s. The sampling interval has been set to 0.1 s. Table \ref{T2} displays the mean absolute percentage error (MAPE) of the prediction results from the three network models.
\begin{table}
  \centering
  \caption{The MAPE of the prediction results.\label{T2}}
  \begin{tabularx}{\linewidth}{Xlll}
    \hline
                  & MLP & RBF & Elman \\
    \hline
    Engine torque                      & 0.81\% & 0.49\% & 1.44\%        \\
    Engine speed         & 1.08\%  & 0.20\% & 1.62\%        \\
   AFR     & 0.95\%  & 1.13\%  & 1.92\%       \\
    \hline
  \end{tabularx}
\end{table}

The comparison results reveal that the prediction of the network models correspond well to the engine output during the model validation phase, where the maximum MAPE is 1.92\%. The RBF network model has the highest accuracy with the MAPEs of 0.49\%, 0.20\% and 1.13\%. This is followed by the MLP model and the Elman model, in order of accuracy. Due to the accuracy and robustness advantages of the RBF network model, it will be used in the development of the AMPC in the following section.

\section{RBF-based adaptive model predictive control}
Adapting the prediction model to changing operating conditions allows the AMPC to address highly nonlinear control
problems. As previously mentioned, updating the prediction model states with an off-line LPV model is expected to enable efficient and robust AMPC strategies. 
Consequently, this section establishes an LPV model based on the parameters of the previously selected RBF network model, facilitating the development of AMPC for the DFLS. Specifically, the AMPC system structure is introduced first, followed by the presentation of the LPV model constructed through a network associated with the RBF model. Finally, simulation studies on the RBF-based AMPC for DFLS control are conducted.

\subsection{Control system structure}
\begin{figure*}[h]
\centering                             
\includegraphics[width=1\linewidth]{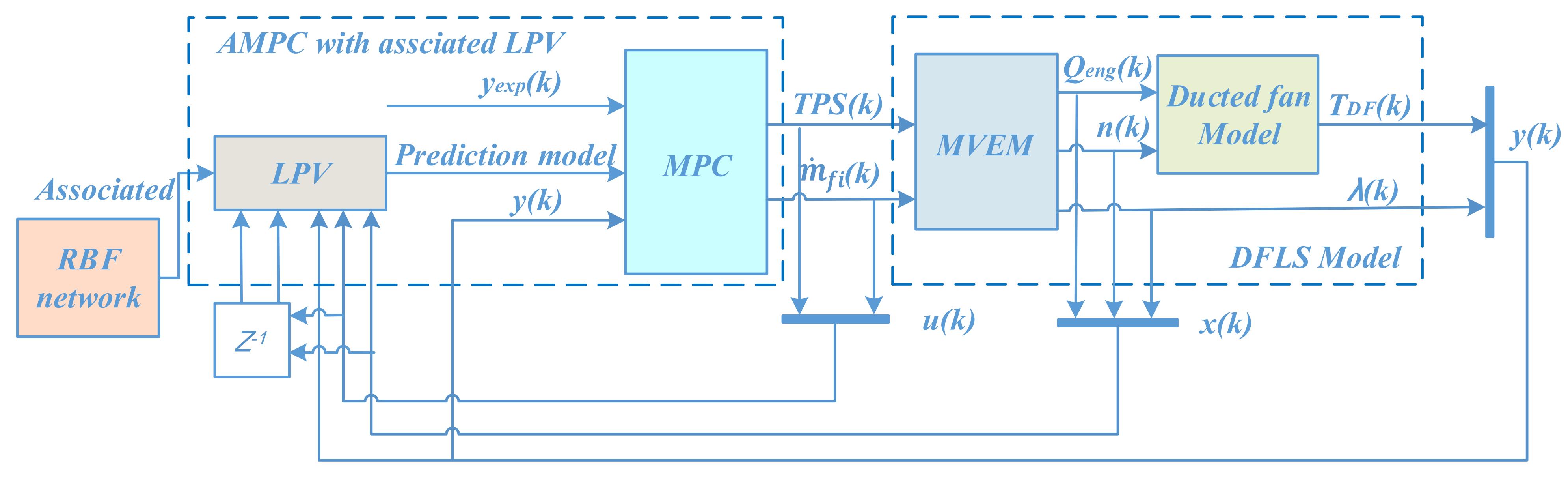} 
\caption{The block diagram of the AMPC structure.}\label{fig:22}
\end{figure*}

The idea of adaptive model predictive control has been introduced in details in the literature\cite{2010Online}. Figure \ref{fig:22} shows the block diagram of the AMPC structure. The nonlinear DFLS model is consisting of the MVEM and the ducted fan modules in the control simulations. At each time step, the LPV model associated with the RBF network will generate a constant linear prediction model based on the current state and input.

The tracking control target is to minimize the error between the system output $\bm{y}=[T_{DF},\lambda]^T$ and the reference $\bm{y}^{ref}$ over a specified time horizon. The AMPC controller will be used to optimize two DFLS inputs, $\bm{u}=[TPS,\dot{m}_{fi}]^T$, to achieve precise dynamic tracking of the desired DFLS thrust ${T_{DF}}^{ref}$ and the engine AFR $\lambda^{ref}$. A cost function $Z$ is therefore defined as a quadratic function in terms of tracking errors and input increments: 

\begin{equation}
\label{eq:16}
\begin{aligned}
Z(k) &= \epsilon\sum_{j=k+N_1}^{k+N_2}\{[\lambda^{ref}(j)-\hat{\lambda}(j)]^2+[{T_{DF}}^{ref}(j)-\hat{T}_{DF}(j)]^2\}\\
&+\xi\sum_{j=k}^{k+N_c}\{[\dot{m}_{fi}(j)-\dot{m}_{fi}(j-1)]^2+[TPS(j)-TPS(j-1)]^2\},
\end{aligned}
\end{equation}
where $k$ represents the current time instance, $\hat{\lambda}$ is the predicted AFR, $\hat{T}_{DF}$ is the predicted thrust of the DFLS, and $\epsilon$ and $\xi$ are control weighting factors that penalize the excessive modification of control inputs. Future horizon $N_1$ and $N_2$, also known as the prediction horizon, specifies the number of upcoming samples. Likewise, the control horizon $N_c$ specifies the number of samples from which optimal inputs are calculated. At each sampling instant, the optimization yields a sequence of input signals, but only the first input is applied to the plant\cite{Alessio2009, Jay2011Model} resulting in a receding horizon approach\cite{2006Adaptive}.

\subsection{The LPV model}

The estimation of the state and the update of the prediction model are indispensable to the effectiveness, precision, and robustness of an AMPC\cite{2009Neural}. The AMPC has currently adopted three categories of model updating strategies: online parameter estimation, successive linearization, and linear parameter varying (LPV) modeling. 
The online parameter estimation method estimates model parameters from real-time plant measurements and updates the prediction model. However, it can be computationally intensive, and suitable for control applications with longer control intervals and sufficient resources. 
The successive linearization method builds the plant model using nonlinear equations and derives a linear approximation to update the model parameters. But dealing with highly nonlinear systems using successive linearization often requires many iterations and computations. 
An LPV system is a linear state-space model whose dynamics vary as a function of certain time-varying parameters. Using an off-line LPV model has the advantage of supporting rapid batch linearization to obtain a variety of plant models at the desired operating points for model updating. This is advantageous for the computation efficiency and stability of the DFLS AMPC. However, LPV models of nonlinear mechanical systems are typically constructed based on the system's dynamic model, which is challenging for SI engines. 

This paper proposes to directly utilize the trained RBF network to construct the LPV model for the AMPC prediction step. Specifically, an associated network can be derived from the RBF model off-line\cite{Zhou2022} and outputs directly the Jacobian matrix of the RBF model, which contains the LPV model parameters. 
The LPV model of the DFLS can be expressed in the discrete linear form with time-varying parameters:
\begin{equation} \label{eq:17}
\Delta \bm{x}(t+1) = A(t) \Delta \bm{x}(t)+B(t) \Delta \bm{u}(t),
\end{equation}
\begin{equation} \label{eq:18}
\Delta \bm{y}(t) = C(t) \Delta \bm{x}(t)+D(t) \Delta \bm{u}(t),
\end{equation}
where the states, inputs, and outputs of the DFLS are
\begin{equation} \label{eq:21}
\bm{x}(t) =[Q_{eng}(t),n(t),\lambda(t)],
\end{equation}
\begin{equation} \label{eq:19}
\bm{u}(t) =[TPS(t),\dot{m}_{fi}(t)], and
\end{equation}
\begin{equation} \label{eq:20}
\bm{y}(t) =[T_{DF}(t),\lambda(t)].
\end{equation}
Because the current state $Q_{eng}(t)$ has no effect on the subsequent system states, the first column of the system matrix $A(t)$ contains only zeros. And the linearized system's matrices can be written as follows:
\begin{equation} \label{eq:22}
A(t) = \frac{\partial \bm{x}(t+1)}{\partial \bm{x}(t)}=\begin{bmatrix}
0 & \frac{\partial Q_{eng}(t+1)}{\partial n(t)} & \frac{\partial Q_{eng}(t+1)}{\partial \lambda(t)}\\
0 & \frac{\partial n(t+1)}{\partial n(t)} & \frac{\partial n(t+1)}{\partial \lambda(t)}\\
0 & \frac{\partial \lambda(t+1)}{\partial n(t)} & \frac{\partial \lambda(t+1)}{\partial \lambda(t)}
\end{bmatrix},
\end{equation}

\begin{equation} \label{eq:23}
B(t) = \frac{\partial \bm{x}(t+1)}{\partial \bm{u}(t)}=\begin{bmatrix}
    \frac{\partial Q_{eng}(t+1)}{\partial TPS(t)} & \frac{\partial Q_{eng}(t+1)}{\partial \dot{m}_{fi}(t)}\\
\frac{\partial n(t+1)}{\partial TPS(t)} & \frac{\partial n(t+1)}{\partial \dot{m}_{fi}(t)}\\
\frac{\partial \lambda(t+1)}{\partial TPS(t)} & \frac{\partial \lambda(t+1)}{\partial \dot{m}_{fi}(t)}
\end{bmatrix},
\end{equation}

\begin{equation} \label{eq:24}
C(t) = \frac{\partial \bm{y}(t+1)}{\partial \bm{x}(t+1)}=\begin{bmatrix}
\frac{\partial T_{DF}(t+1)}{\partial Q_{eng}(t+1)} & \frac{\partial T_{DF}(t+1)}{\partial n(t+1)} & 0\\
0 & 0 & 1
\end{bmatrix},
\end{equation}

\begin{equation} \label{eq:25}
D(t) = \frac{\partial \bm{y}(t+1)}{\partial \bm{u}(t+1)}=\begin{bmatrix}
0 & 0\\
0 & 0 
\end{bmatrix}.
\end{equation}
The output matrix $C$ can be calculated explicitly using equations \ref{eq:4}-\ref{eq:7} from the ducted fan model. And the system matrix $A$ and the control effectiveness matrix $B$ are associated with the SI engine, the elements of which can be approximated by the partial derivatives of the trained RBF network inputs $\textbf{\textit{h}} =[TPS(t),\dot{m}_{fi}(t), n(t),\lambda(t)]$ with respect to the network outputs $\textbf{\textit{p}}=[Q_{eng}(t+1),n(t+1),\lambda(t+1)]$. 

To determine the Jacobian matrix of the RBF model $\mathcal{J}$, we can rewrite the $j$th Gaussian radial function in the RBF network from equation \ref{eq:15.1} as below:
\begin{equation} \label{eq:15.2}
\phi_j(p) =e^{-({\|\bm{p}-\bm{c}_j\|}/{{s_j}})^2}.
\end{equation}
where $c_j$ is the $j$th center point and $s_j$ is its radius. The associated network is linear to the parameters $LW$, which are weights connecting the output of each radial function to the output of the network, when $c_j$ and $s_j$ are fixed. 
Derivatives can therefore be explicitly calculated as a function of the network input $\textbf{\textit{p}}$ and network weights $LW$:
\begin{equation} \label{eq:15.3}
\mathcal{J}(\textbf{\textit{p}},LW) =LW^T\Phi(\textbf{\textit{p}})=LW^T\begin{bmatrix}
\phi_1(\textbf{\textit{p}})\frac{-2}{{s_1}^2}(\textbf{\textit{p}}-\textbf{\textit{c}}_1) \\
\vdots\\
\phi_J(\textbf{\textit{p}})\frac{-2}{{s_J}^2}(\textbf{\textit{p}}-\textbf{\textit{c}}_J) 
\end{bmatrix}.
\end{equation} 
And this function can be constructed as another network associated with the RBF model network as depicted in Figure \ref{fig:111}, with $J$ activation functions, each of which is the partial derivative of the $j$th Gaussian radial function output with respect to the critic input:
\begin{figure}[ht]
\centering                             
\includegraphics[width=0.55\linewidth]{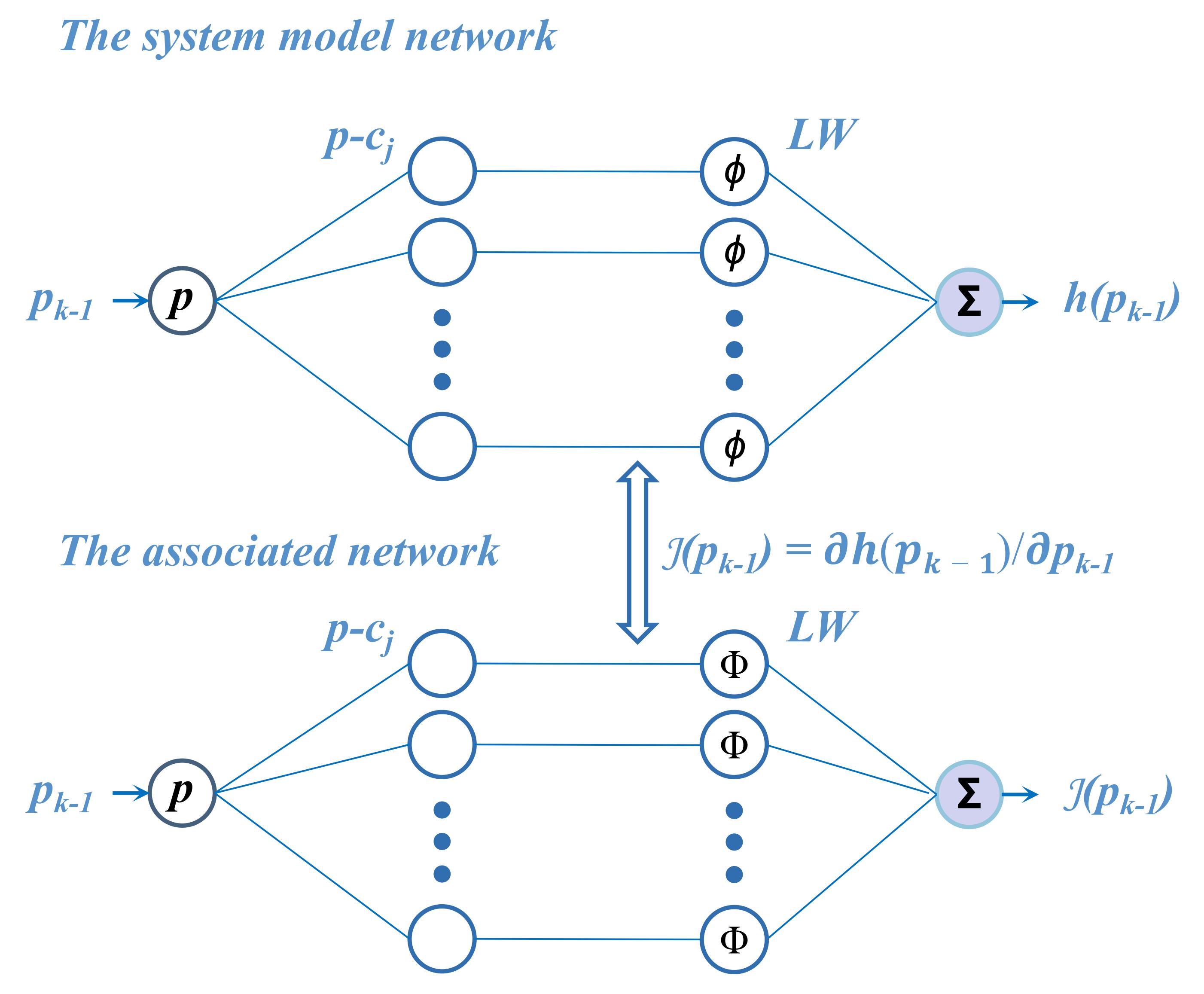} 
\caption{The system model network with RBF activation functions and its associated
network. These two networks share the same set of centers, denoted by $c_j$, and the to-be-determined parameters $LW$.\cite{Zhou2022}}\label{fig:111}
\end{figure}

\begin{equation} \label{eq:15.4}
\begin{aligned}
\Phi_j(\bm{p}) &= \frac{\phi_j(\bm{p})}{\partial \bm{p}} = e^{-({\|\bm{p}-\bm{c}_j\|}/{{s_j}})^2}\frac{-2}{{s_j}^2}(\bm{p}-\bm{c}_j)\\&=\phi_j(\bm{p})\frac{-2}{{s_j}^2}(\bm{p}-\bm{c}_j),
\end{aligned}
\end{equation}
where $\Phi_j$ represents the $j$th row vector of the derivatives $\Phi(\bm{p}) = \partial \phi(\bm{p})/\partial \bm{p}$.

This associated network model outputs directly the Jacobian matrix of the RBF model $\mathcal{J}$, whose elements are the partial derivatives of the model outputs with respect to the model inputs, and which can be used to construct the LPV model matrices $A$ and $B$ in equations \ref{eq:22} and \ref{eq:23}. 
Note that this concise mathematical relationship is derived from the linear-in-parameter property of the system model \cite{Zhou2022}, which will effectively enhance the AMPC prediction model's update speed. Moreover, employing a linear-in-parameter system model can aid in avoiding the local minimum trap. 
The system model network is intended to be nonlinear in inputs but linear in parameters, and the associated network can be derived explicitly and precisely from the system model network. This characteristic is indispensable for the computational efficiency of the RBF-based AMPC.

\subsection{AMPC simulation of the DFLS}
To validate the proposed AMPC, the control simulation of the DFLS during takeoff thrust preparation is implemented. During the vertical take-off power preparation procedure, the anticipated thrust of the DFLS ${T_{DF}}^{ref}$ gradually increases from the idling rating (10 $kgf$) to the hovering thrust (80 $kgf$). During this procedure, the engine gradually increases its throttle to a relatively stable state with a low expected normalized AFR (0.82), ensuring adequate power output, before adjusting the expected normalized AFR (1.0) to an efficient mode. In the simulation study, the throttle position is constrained between 5\% and 90\%, and the fuel injection rate is between 0.0011 and 0.0055 kg/s. The thrust of the DFLS is constrained between 0 and 150 kgf, and the normalized AFR is between 0.68 and 1.26.

\begin{figure}[h] 
\centering
\includegraphics[width=0.66\linewidth]{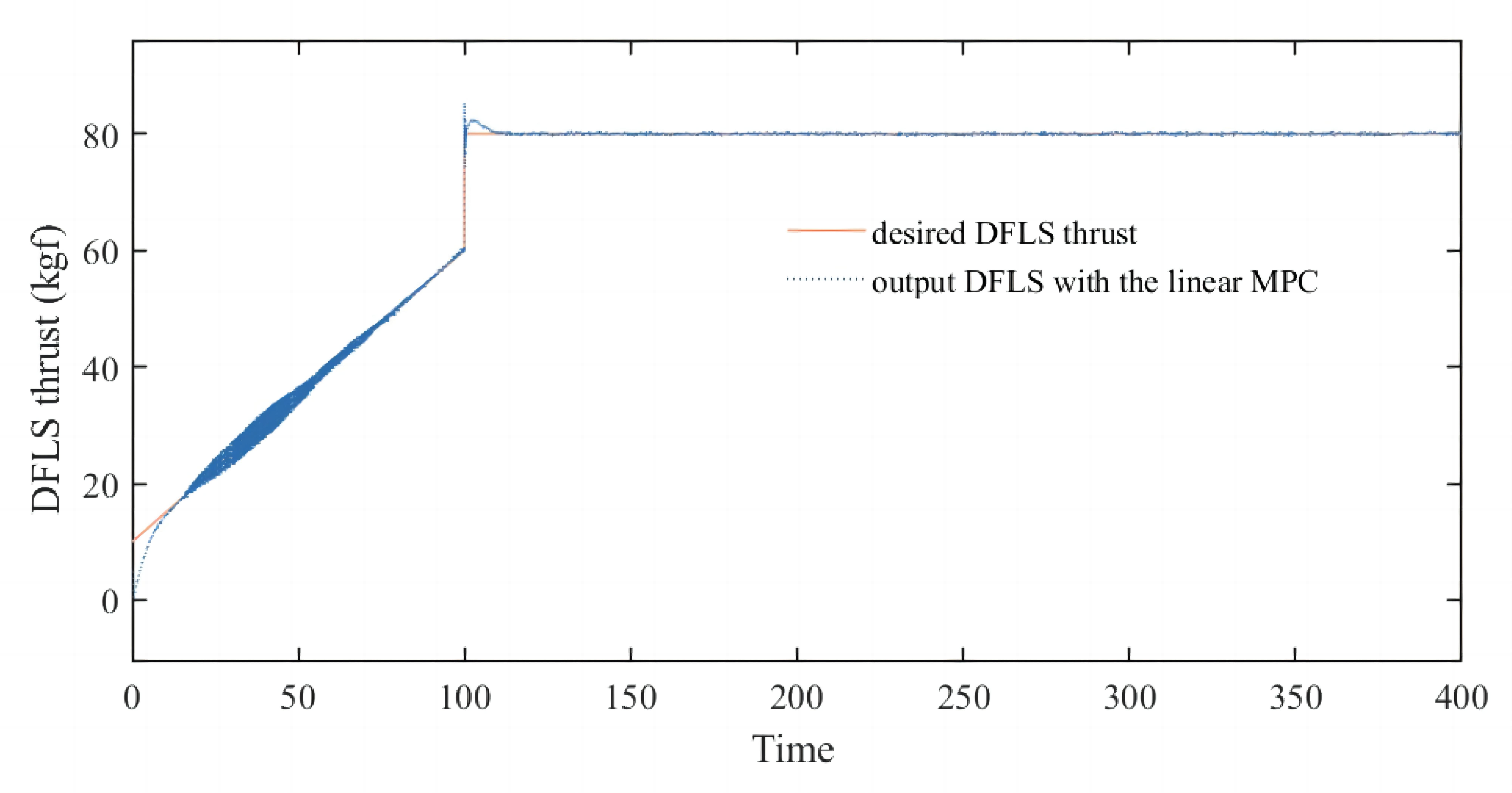} 
\caption{The DFLS thrust tracking performance with a linear MPC.}\label{fig:31}
\end{figure}

\begin{figure}[h]  
\centering
\includegraphics[width=0.66\linewidth]{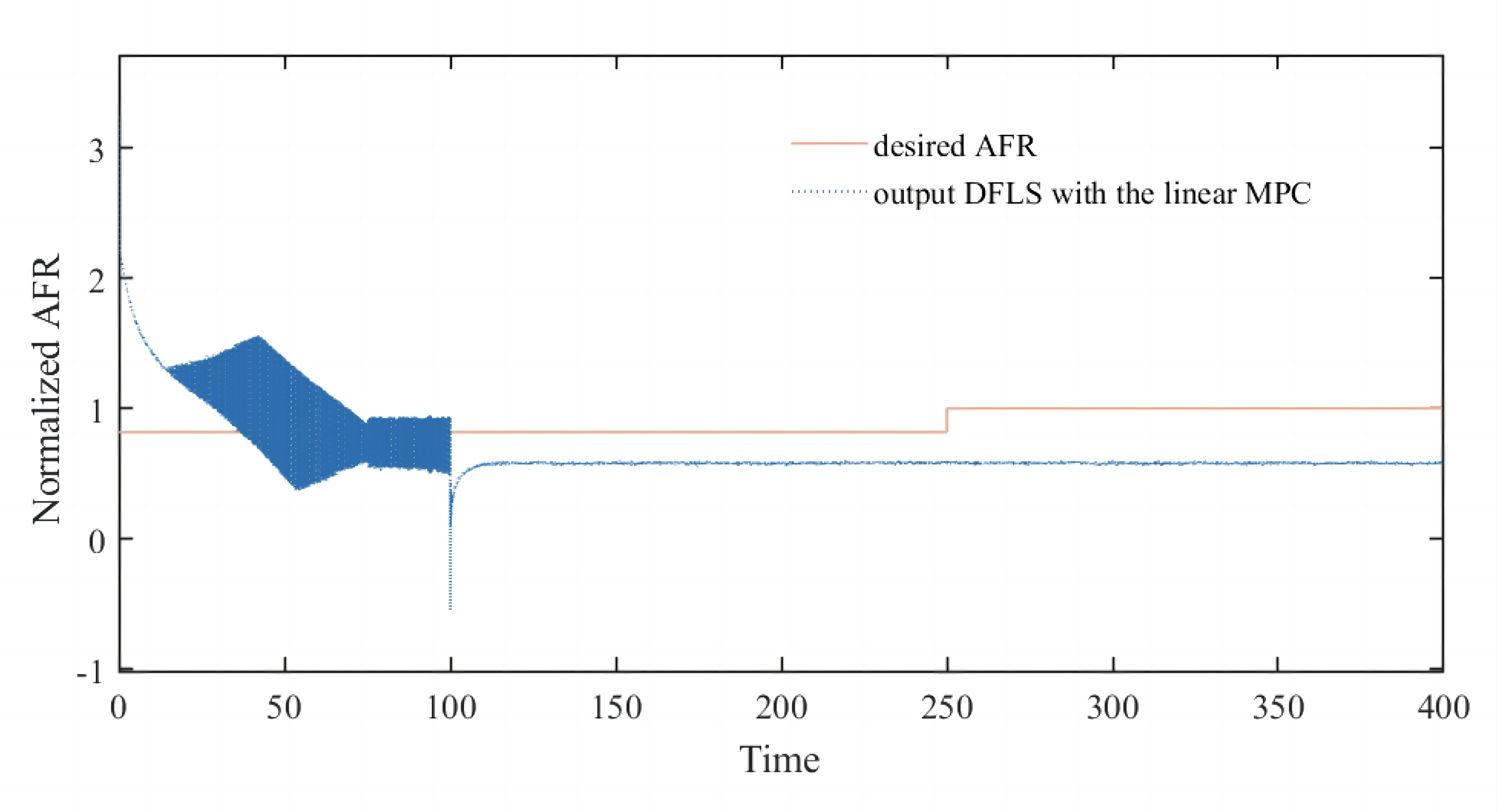} 
\caption{The AFR tracking performance with a linear MPC.}\label{fig:32}
\end{figure}

In order to compare the AMPC against a traditional MPC, a linear MPC is designed for the same control process for the DFLS. The linear MPC and the AMPC are designed with the same cost function, as illustrated in equation \ref{eq:16}, and same parameters. The nonlinear optimization parameters were set as follows: $N_1 = 1$, $N_2 = 8$, $N_c = 3$, $\epsilon=0.8$, and $\xi=0.5$. In the control simulation, the normalized AFR measurement and the DFLS thrust measurement were subjected to Gaussian noise $\mathcal{N}$(0, 0.005). Figures \ref{fig:31}-\ref{fig:32} depict the simulation tracking results for the DFLS's desired thrust and AFR. Before time step $100$, the desired thrust is continuously increasing, the tracing control results of the DFLS thrust and the engine AFR with the linear MPC are divergent. 
After the expected thrust has been stabilized, the linear MPC can only precisely track the expected thrust, while the engine AFR control result is inconsistent with the desired value. As expected, the linear MPC's tracking and regulating performances are unacceptable.

\begin{figure}[htbp]
\centering
{
    \begin{minipage}[t]{1\linewidth}
        \centering
        \includegraphics[scale=0.057]{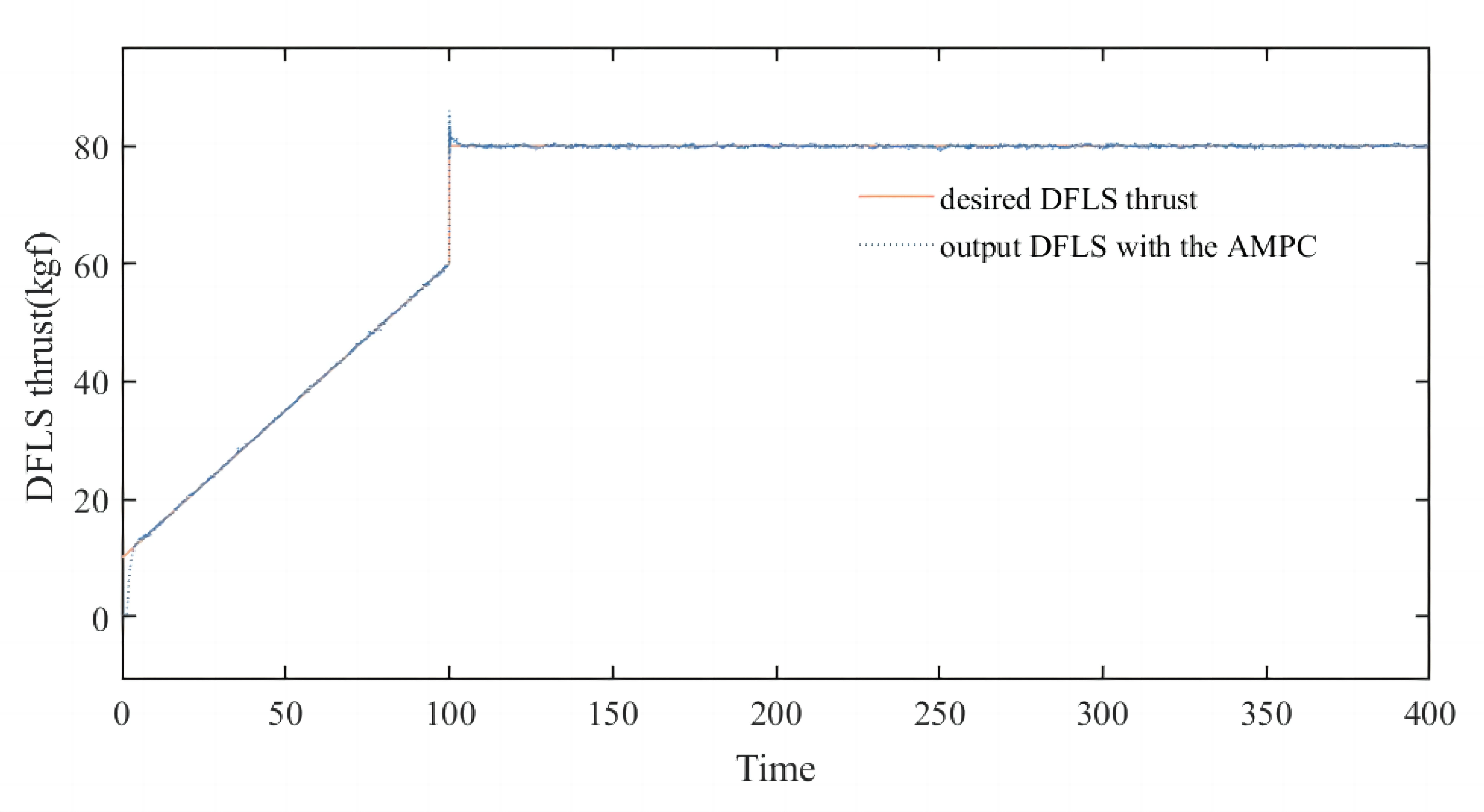} \\{\fontsize{8pt}{1em} \selectfont a) Tracking performance of the desired thrust of the DFLS \par}
    \end{minipage}
}
{
 	\begin{minipage}[t]{1\linewidth}
        \centering
        \includegraphics[scale=0.057]{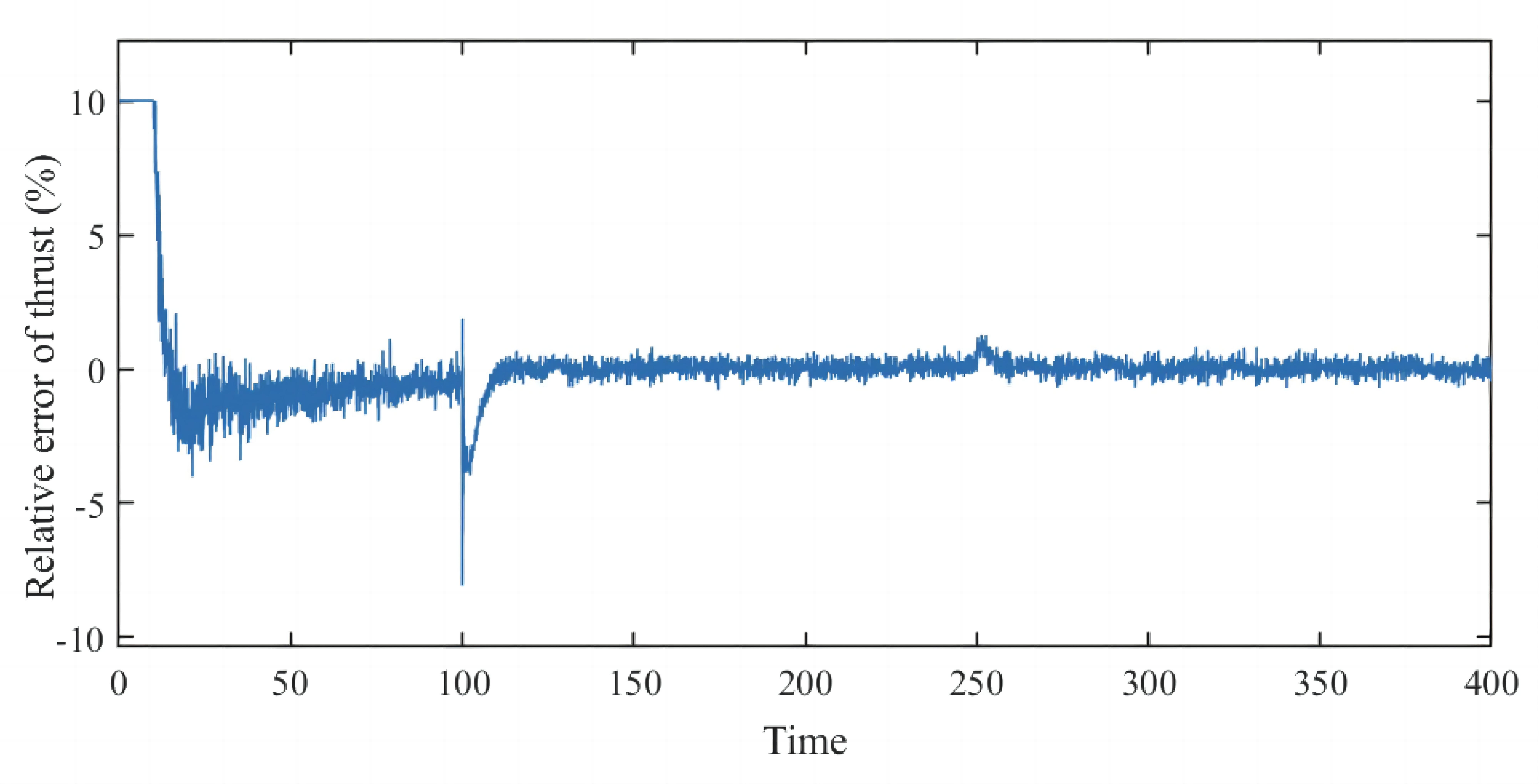} \\{\fontsize{8pt}{1em} \selectfont b) Relative error of the thrust control \par}
    \end{minipage}
}
\caption{The DFLS thrust tracking performance with the proposed AMPC.}\label{fig:23}
\end{figure}

\begin{figure}[htbp]
\centering
{
    \begin{minipage}[t]{1\linewidth}
        \centering
        \includegraphics[scale=0.056]{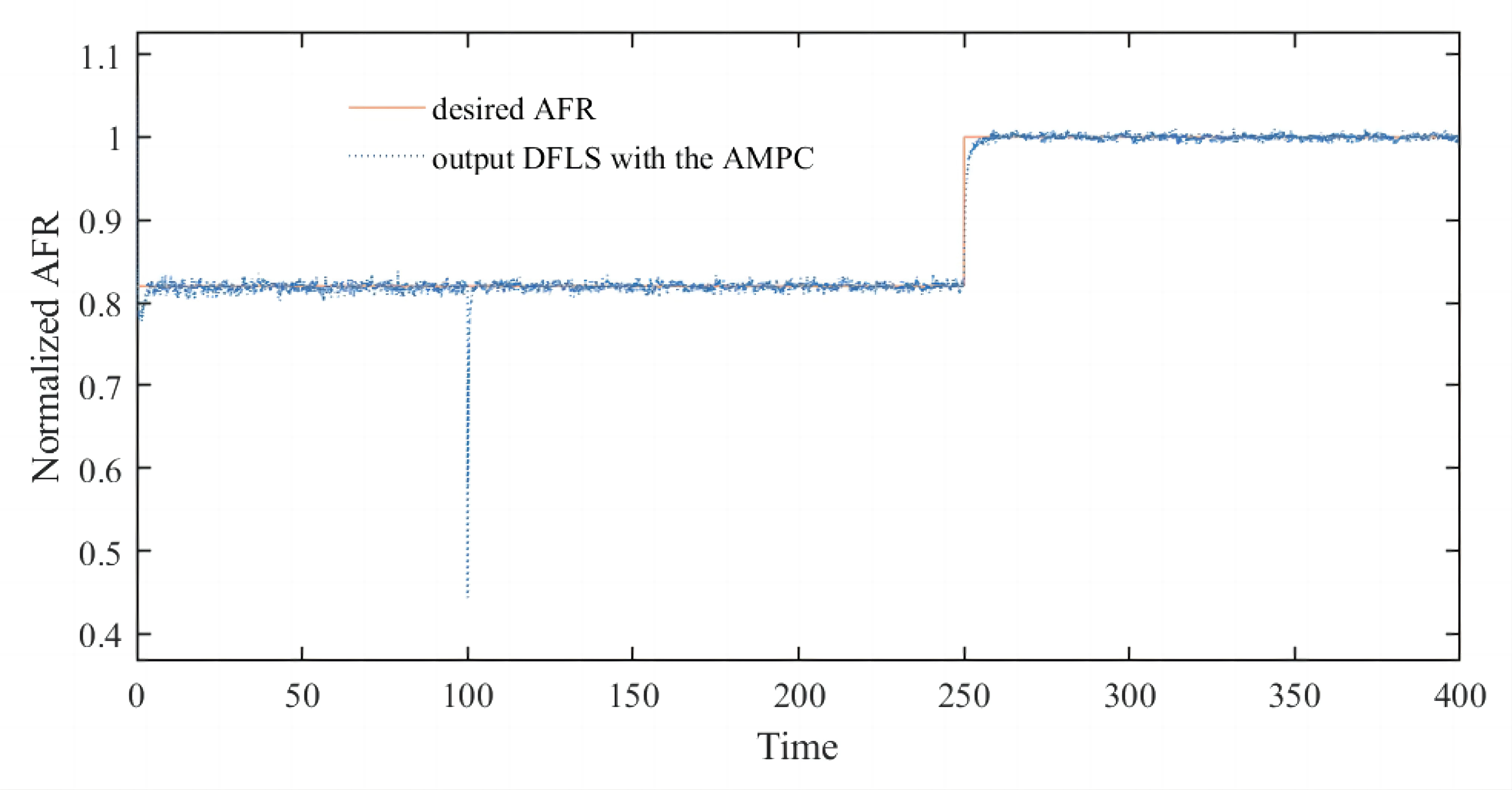} \\{\fontsize{8pt}{1em} \selectfont a) Tracking performance of the desired AFR \par}
    \end{minipage}
}
{
 	\begin{minipage}[t]{1\linewidth}
        \centering
        \includegraphics[scale=0.057]{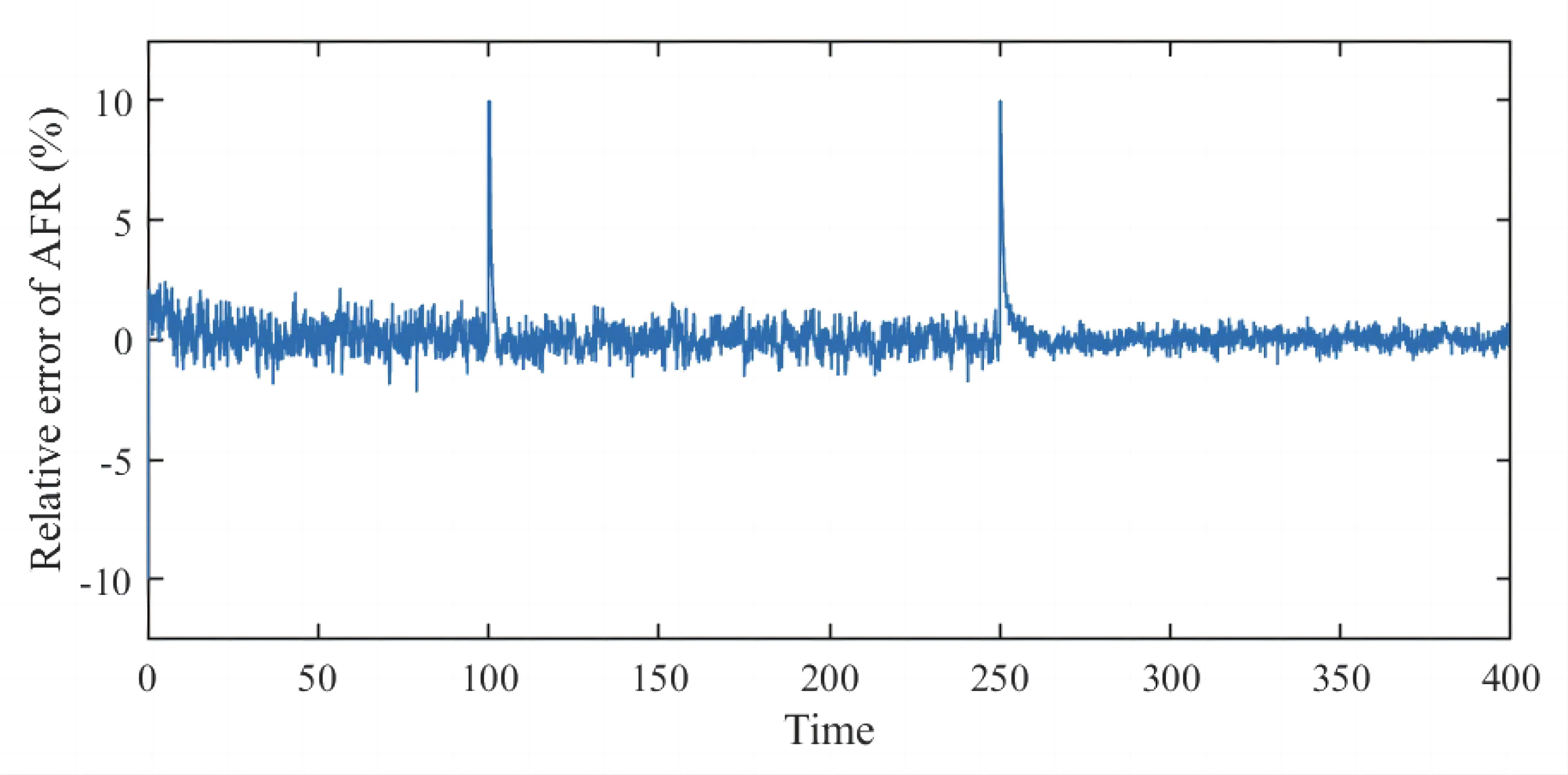} \\{\fontsize{8pt}{1em} \selectfont b) Relative error of the AFR control \par}
    \end{minipage}
}
\caption{The AFR tracking performance with the proposed AMPC.}\label{fig:24}
\end{figure}

\begin{figure}[htbp]
\centering
{
    \begin{minipage}[t]{1\linewidth}
        \centering
        \includegraphics[scale=0.5]{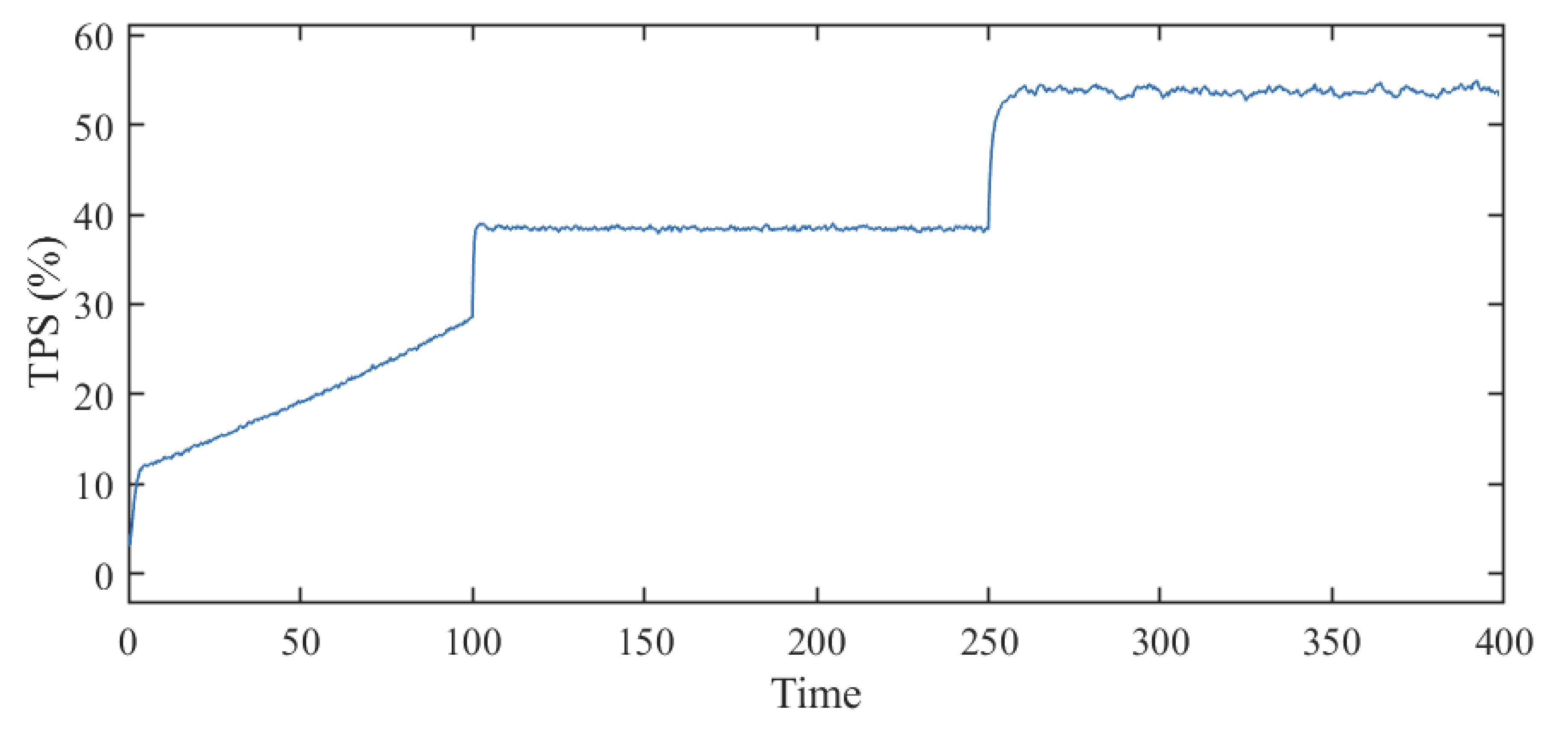} \\{\fontsize{8pt}{1em} \selectfont a) Optimized throttle position by the AMPC \par}
    \end{minipage}
}
{
 	\begin{minipage}[t]{1\linewidth}
        \centering
        \includegraphics[scale=0.056]{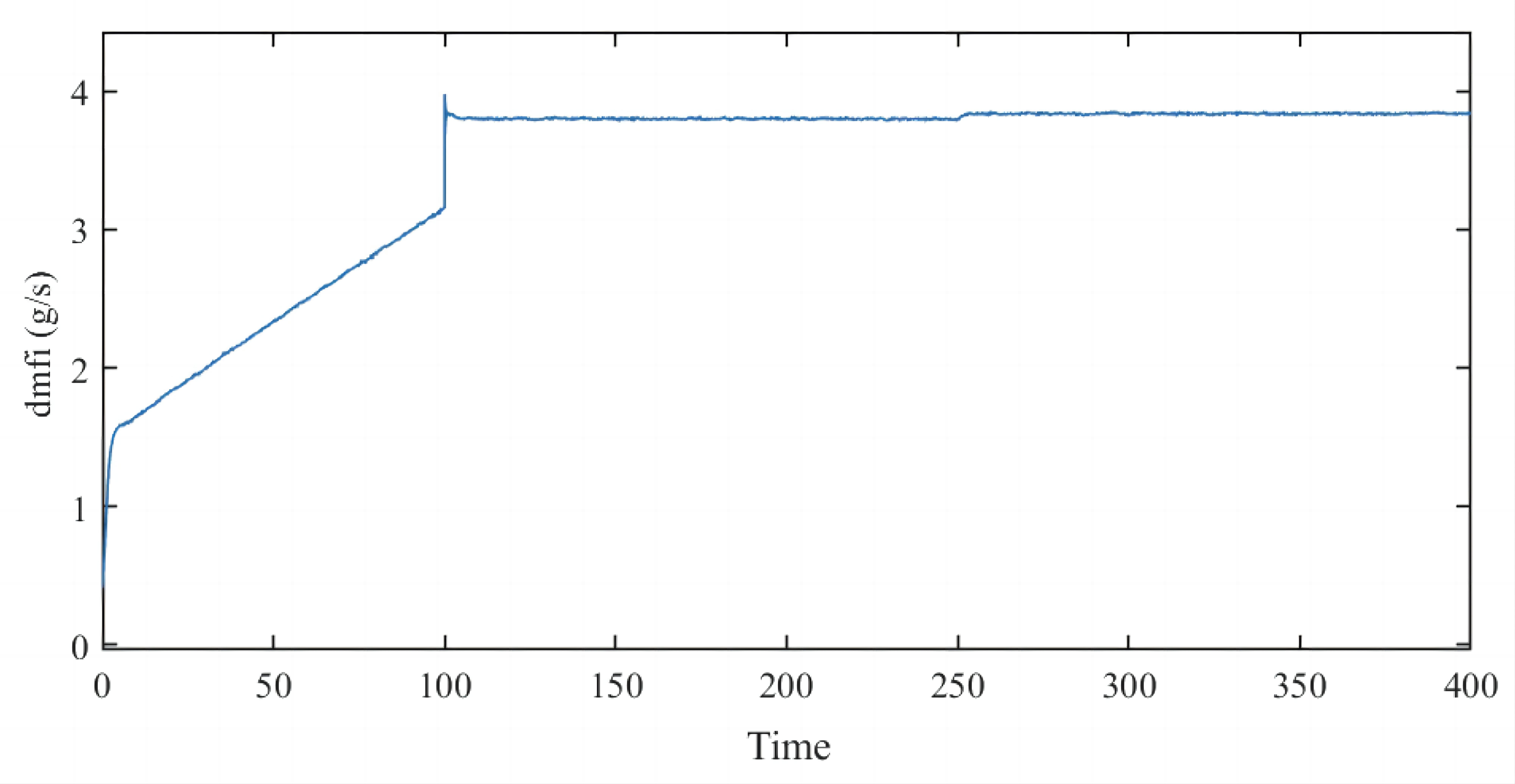} \\{\fontsize{8pt}{1em} \selectfont b)  Optimized fuel injection mass flow by the AMPC \par}
    \end{minipage}
}
\caption{The control inputs of the engine (the TPS and the fuel injection mass flow $\dot{m}_{fi}$) optimized by the proposed AMPC.}\label{fig:25}
\end{figure}

\begin{figure}[htbp]
\centering
{
    \begin{minipage}[t]{1\linewidth}
        \centering
        \includegraphics[scale=0.057]{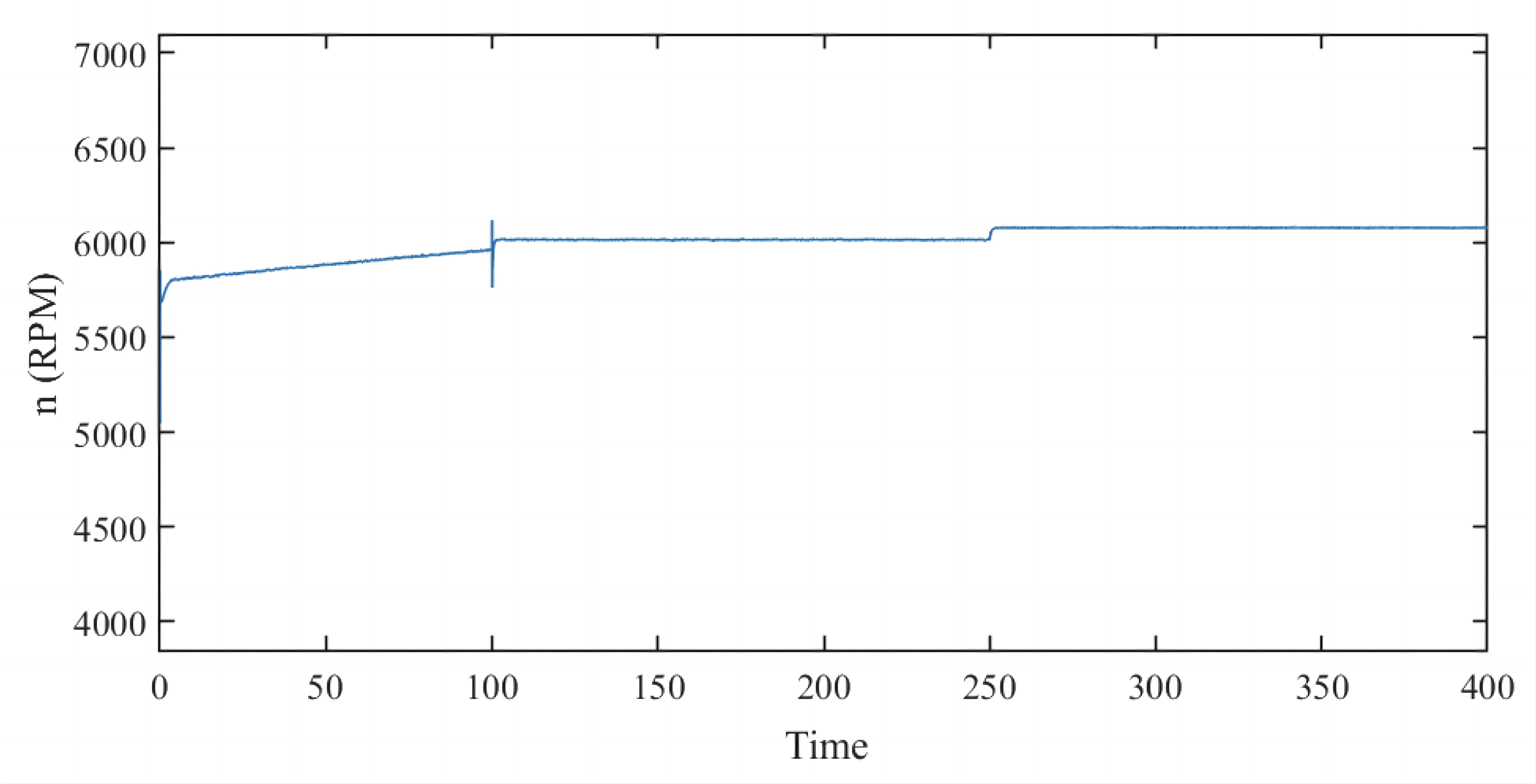} \\{\fontsize{8pt}{1em} \selectfont a) The engine crankshaft speed  \par}
    \end{minipage}
}
{
 	\begin{minipage}[t]{1\linewidth}
        \centering
        \includegraphics[scale=0.057]{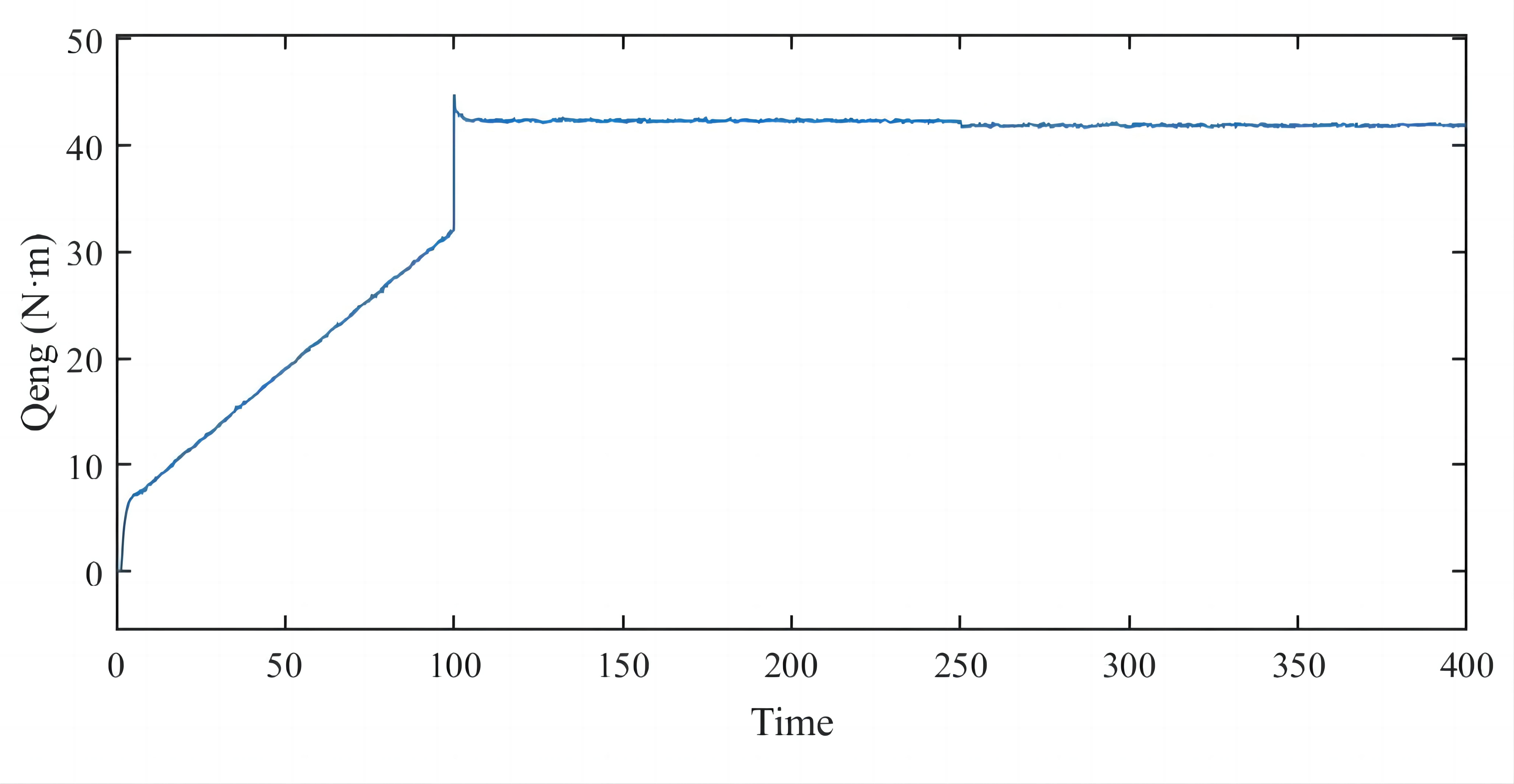} \\{\fontsize{8pt}{1em} \selectfont b) The engine crankshaft torque \par}
    \end{minipage}
}
\caption{The engine crankshaft speed $n$ and the engine crankshaft torque $Q_{eng}$ during the control simulation.}\label{fig:26}
\end{figure}

The AMPC parameters are manually adjusted to determine the appropriate values by comparing the control performance. Figure \ref{fig:23} depicts the tracking control results for the desired thrust. Figure \ref{fig:24} depicts the corresponding simulation results of the tracking curve for the desired normalized AFR. Figure \ref{fig:25} illustrates the AMPC-optimized $TPS$ and $\dot{m}_{fi}$ engine results, respectively. Figure \ref{fig:26} depicts the corresponding simulation results for the engine speed and output torque. 


\newpage
The simulation results indicate that the developed AMPC can accurately control the thrust and AFR during the vertical take-off power preparation process. As illustrated in Figures \ref{fig:23},\ref{fig:24}, the relative error ranges for DFLS thrust and SI engine AFR tracking control are -3.5\% to 1.0\% and -2.1\% to 2.2\%, respectively. The aforementioned outcomes validated the effectiveness and robustness of the proposed RBF-based AMPC.

\section{Conclusions}
This paper presents a novel adaptive model predictive control (AMPC) approach for controlling the thrust of an engine-driven ducted fan lift system (DFLS). 
The proposed method is based on an off-line linear parameter varying (LPV) model, whose parameter updating laws is derived from a radial basis function (RBF) network. 
The use of the RBF network is motivated by its superior prediction accuracy and robustness compared to other network models, such as multi-Layer perceptron and Elman. 
This LPV model enables real-time updates of the AMPC prediction model across the full operating envelope, making it an effective solution for handling the highly nonlinear dynamics of DFLS. 
The proposed AMPC receives and updates its LPV parameters from an associated network without massive online operations, which enhances the control effectiveness and avoids model errors resulting from issues such as delays, noise, and insufficient excitation. 
This concise mathematical relationship is derived from the linear-in-parameter property of the system model, avoiding local minimum traps. 
The DFLS AMPC was validated in numerical simulations, demonstrating its ability to achieve precise control of thrust and air-fuel ratio during the vertical take-off preparation process. 
The control strategy is designed to track the desired thrust by controlling the engine output power while ensuring reliability and efficiency through synchronized air-fuel ratio control.

The RBF model-based AMPC approach proposed in this paper is efficient and practical. The validation results show its potential for wider application to other nonlinear industrial control problems. However, the LPV model obtained off-line may lack sufficient robustness to handle stochastic and uncertain plant model changes. Further research is needed to explore the combination of the proposed method with online AMPC to preserve control efficiency and enhance its adaptability. In conclusion, this study presents a promising new method for controlling the thrust of engine-driven ducted fan lift systems and opens up avenues for further improvement and refinement.

\section*{Acknowledgement} 

The corresponding author would like to thank Malaysian Ministry of Higher Education (MOHE) for providing the Fundamental Research Grant Scheme (FRGS): FRGS/1/2020/TK0/USM/03/11.






\bibliographystyle{unsrt}
\bibliography{example}






\end{document}